%% file: NM_emulator_prc.tex
\documentclass[aps,prc,twocolumn,superscriptaddress,longbibliography]{revtex4-1}
\usepackage{epsfig,dsfont,amssymb,amsmath,amsthm,amsfonts,amsbsy,mathrsfs}
\usepackage{graphicx}
\usepackage{amsmath}
\usepackage{amssymb}%
\usepackage{dcolumn}
\usepackage{hyperref}
\usepackage[color=green!30]{todonotes}

\newcommand{\be}{\begin{equation}}
\newcommand{\ee}{\end{equation}}
\newcommand{\ba}{\begin{eqnarray}}
\newcommand{\ea}{\end{eqnarray}}

\input{./shorthands.tex}

\begin{document}

\title{Emulating \emph{ab initio} computations of infinite nucleonic matter}

\author{W.\ G.\ Jiang}
\affiliation{Department of Physics, Chalmers University of Technology, SE-412 96 G\"oteborg, Sweden}
\affiliation{Institut f\"ur Kernphysik and PRISMA Cluster of Excellence, Johannes Gutenberg Universit\"at, 55128 Mainz, Germany}

\author{C.\ Forss\'en}
\affiliation{Department of Physics, Chalmers University of Technology, SE-412 96 G\"oteborg, Sweden}

\author{T.\ Dj\"arv}
\affiliation{Department of Physics, Chalmers University of Technology, SE-412 96 G\"oteborg, Sweden}
\affiliation{Physics Division, Oak Ridge National Laboratory, Oak Ridge, TN 37831, USA}


\author{G.\ Hagen}
\affiliation{Physics Division, Oak Ridge National Laboratory, Oak Ridge, TN 37831, USA}
\affiliation{Department of Physics and Astronomy, University of Tennessee, Knoxville, TN 37996, USA}

\begin{abstract}
We construct efficient emulators for the \emph{ab initio} computation of the infinite nuclear matter equation of state. These emulators are based on the subspace-projected coupled-cluster method for which we here develop a new algorithm called small-batch voting to eliminate spurious states that might appear when emulating quantum many-body methods based on a non-Hermitian Hamiltonian. The efficiency and accuracy of these emulators facilitate a rigorous statistical analysis within which we explore nuclear matter predictions for $> 10^6$ different parametrizations of a chiral interaction model with explicit $\Delta$-isobars at next-to-next-to leading order.
Constrained by nucleon-nucleon scattering phase shifts and bound-state observables of light nuclei up to \nuc{4}{He}, we use history matching to identify non-implausible domains for the low-energy coupling constants of the chiral interaction.
Within these domains we perform a Bayesian analysis using sampling/importance resampling with different likelihood calibrations and study correlations between interaction parameters, calibration observables in light nuclei, and nuclear matter saturation properties.
\end{abstract}

\maketitle
\section{Introduction}
Infinite nuclear matter is an idealized system of strongly interacting nucleons that holds translational invariance without Coulomb and surface effects. Studies of its \EOS{} at nuclear densities allow to explore properties of the microscopic interactions between constituent nucleons and to understand bulk properties of finite nuclei.
It also provides an important anchor point for extrapolations to higher densities which is needed for the description of neutron stars and their mergers~\cite{hebeler2015,lynn2019,Leonhardt:2019fua,Greif:2020pju,drischler2021,huth2021,Schatz:2022vzq}.
While previous theoretical studies are mostly based on fixed parametrizations of certain interaction models~\cite{song1998,carlson2003,gandolfi2009,hebeler2011,rios2011,baldo2012,baldo2012,carbone2013,hagen2013b,baardsen2013,tews2015,afanasjev2016,bombaci2018,carbone2018,drischler2019,drischler2020a,drischler2020b} a rigorous statistical analysis will require to incorporate all relevant sources of uncertainty including the parametric one. In this work we introduce several key developments that are needed to perform such an analysis and we study in particular the multi-dimensional parameter domain of $\Delta$-full \chiEFT{} at \NNLO{}. The sensitivity of nuclear matter predictions to the calibration observables is explored in a companion paper~\cite{jiang2022:short}.

In recent years, the studies of nuclear matter have proved to be informative in various fields. For instance, the \EOS{} of \PNM{} is critical to the astrophysics of supernova explosion \cite{shen1998,burrows2013} and neutron star properties \cite{heiselberg2000b,lattimer2000,lattimer2007,hebeler2013b, hebeler2015, bombaci2018, lynn2019,drischler2021,huth2021} while the incompressibility of \SNM{} is connected with the giant monopole resonance \cite{shlomo1993,youngblood1999,piekarewicz2002}. Furthermore, saturation properties (symmetry energy and saturation density) have been shown to be correlated with selected observables in finite nuclei~\cite{brown2000, furnsthal2002,rocamaza2011,tsang2012,abrahamyan2012, reinhard2013, afanasjev2016} and may help constrain \NN{} plus \TNF[s]~\cite{lejeune2001,hebeler2011,dutra2012,tews2016, jiang2020}. Recently, it was found that the symmetry energy and its slope correlates with the neutron skin and dipole polarizability of the heavy nucleus $^{208}$Pb starting from chiral interactions at \NNLO{} with explicit delta isobars ($\Delta$)~\cite{Hu:2021trw}.

The computational modeling of nuclear matter systems involve several challenges such as finite size effects, shell oscillations, slow convergence and high computational cost. Different many-body methods have been developed to address these problems. Theoretical approaches based on Brueckner-Bethe-Goldstone theory \cite{brueckner1955} have long been used to calculate the \EOS{} \cite{song1998,baldo2012,bombaci2018}. More recently, there has been rapid development of density functional theory (DFT) \cite{erler2012b,afanasjev2016} and relativistic mean-field approaches~\cite{fattoyev2010,rocamaza2011} as well as many-body methods such as many-body perturbation theory~\cite{hebeler2015,drischler2021}, self consistent Green’s functions \cite{rios2011,baldo2012,carbone2013}, Quantum Monte Carlo~\cite{carlson2003,gandolfi2009,tews2015}, and \CC{} methods~\cite{hagen2013b, baardsen2013}. Meanwhile, there has also been great progress in constructing realistic nuclear Hamiltonians based on \chiEFT{} with improved saturation properties~\cite{machleidt2011,hebeler2011,ekstrom2013,ekstrom2015a,jiang2020,soma2020}.

Recent calculations have shown that an accurate description of bulk properties of finite nuclei and nuclear matter involves fine-tuning of the underlying nuclear Hamiltonian~\cite{hebeler2011, ekstrom2015a,carbone2018,simonis2017,morris2018,drischler2019}. The fact that these nuclear Hamiltonians give similar results for scattering phase shifts and few-body observables, but differ for many-body systems.  indicates the challenge we face when constructing nuclear forces. In principle, microscopic approaches based on nucleonic degrees of freedom and chiral interactions should be able to describe both the two- and few-body sectors, as well as infinite nuclear matter. It is not clear whether going to higher orders in the \EFT{} expansion will resolve the fine-tuning, or whether the free parameters of the chiral Hamiltonian, i.e., the \LEC[s], are not sufficiently constrained by standard fitting observables in the few-nucleon sector. Therefore, the construction of nuclear forces to meet the required precision and accuracy is still an ongoing and complex task. Systematic studies, such as the present work, are needed to better understand how details of the interaction influence properties of nucleonic matter, how predictions for different observables may be correlated, and also to rigorously quantify uncertainties. 

Although significant progress has been made towards quantifying uncertainties of \emph{ab initio} nuclear matter predictions and identifying possible correlations with observables in finite nuclei~\cite{hebeler2011,khan2012,hagen2015,kievsky2018,drischler2019,reinhard2016, carson2018, drischler2020a, drischler2020b, Hu:2021trw, Huth:2021bsp}, a full exploration of the sensitivity to variations of the \LEC[s] has been considered too difficult due to the immense computational cost. 

In this work we address this problem by developing accurate emulators and adapting a robust statistical approach known as history matching~\cite{vernon2010,vernon2014,vernon2018} to explore the entire \LEC{} parameter space. History matching is specifically designed to aid the analysis and calibration of high-dimensional computationally expensive physical models. By monitoring predictions of scattering phase shifts and few-body observables we can iteratively identify the (non-implausible) region of the parameter space that gives results consistent with a set of data. This approach then provides a finite domain in which the probability of finding accurate interaction parametrizations is higher and that makes it feasible to perform rigorous statistical studies even with limited computational resources.
 
The history matching procedure and the statistical analysis require a significant amount of computations. Therefore, emulators---that mimic the outputs of the exact calculations at a fraction of the computational cost---are required to bring this kind of statistical approach into practical use. 
Recently, model reduction methods~\cite{melendez2022} such as \EC{}~\cite{Frame2018,ekstrom2019,konig2020,sarkar2020,furnstahl2020,Hu:2021trw} has proved to be an efficient and accurate approach to emulate the predictions of \emph{ab initio} many-body methods.
In this paper, we generalize the \SPCC{}~\cite{ekstrom2019} to construct emulators for the energy per particle of \PNM{} and \SNM{} at different densities that work for a wide range of \LEC[s].
We will demonstrate that these emulators provide fast and accurate approximations to \CC{} calculations of nuclear matter properties \cite{hagen2013b}. While the \SPCC{} method has been successfully applied to a global sensitivity analysis of bulk properties of \nuc{16}{O} \cite{ekstrom2019}, the challenge of the present work is that the bi-variational \CC{} energy functional~\cite{arponen1983,bartlett2007}, combined with the increased level density in infinite nucleonic matter and large number of \LEC[s], may give rise to spurious states when diagonalizing the subspace-projected non-Hermitian \CC{} Hamiltonian. 
We therefore introduce a new algorithm called small-batch voting to efficiently locate the physical ground state and to identify and eliminate spurious states.
These nuclear matter emulators, along with other emulators for light nuclear systems, are then applied to $1.7\times10^6$ different chiral interactions acquired by a history matching procedure. This allows us to carry out a comprehensive study of correlations between nuclear matter properties and observables in finite nuclei. A subsequent Bayesian analysis is performed by establishing error models for \EFT{} truncations, method uncertainties, and using sampling/importance resampling~\cite{smith:1992aa, Jiang:2022off} to obtain probabilistic distributions of both \LEC{} parameters and posterior predictions. Details of nuclear matter posterior predictive distributions calibrated by different sets of observables are elucidated in a companion paper~\cite{jiang2022:short}.

\section{Method%
\label{sec:method}}
In this work we consider $\Delta$-full \chiEFT{} at \NNLO{}~\cite{vankolck1994,ordonez1994,ordonez1996,piarulli2015,piarulli2016,ekstrom2017, jiang2020}. The explicit inclusion of the $\Delta$-isobar degree of freedom is beneficial since it increases the breakdown scale of the \chiEFT{} and gives a better description of nuclear matter properties~\cite{ekstrom2017,jiang2020,Hu:2021trw}. We use standard nonlocal regulators: $f(p)=\exp[-(p/\Lambda)^{2n}]$ and $f(p,q)=\exp\{-[(p^2+3q^2/4)/\Lambda^2]^n\}$ for the \NN{} and \TNF{} interactions respectively, with $n=4$ and a fixed cutoff $\Lambda = 394$~MeV.
The chiral Hamiltonian of $\Delta$NNLO is parametrized with 17 \LEC[s] which are here represented by the vector $\vec{\alpha}$. Following Refs. \cite{ekstrom2019,konig2020} it can be written as:
\begin{eqnarray}
  \label{eq_H_LECs}
  H(\vec{\alpha}) = h_{0} +  \displaystyle\sum_{i=1}^{N_{\rm{LECs}}=17}  \alpha_i h_i,
\end{eqnarray}
with $h_0 = t_{\rm{kin}}+V_0$. Here $t_{\rm{kin}}$ is the kinetic energy and $V_0$ represents the constant potential term.

One of the most important conclusions of \EC{} is that the trajectory of the eigenvectors as a function of the smoothly varying control parameters of the Hamitlonian (in our case the \LEC[s]) can be well described by a finite-dimensional manifold \cite{Frame2018}. With this statement, the ground-state eigenvector of any target Hamiltonian $H(\vec{\alpha}_{\circledcirc})$ can be well approximated by some linear combinations of ground-state eigenvectors of a finite set of training Hamiltonians $H(\vec{\alpha}_{1}),\cdots, H(\vec{\alpha}_{N_{\rm{sub}}})$. 

In practice, to create a subspace emulator we therefore need to use a many-body solver to generate $N_{\rm{sub}}$ different ground-state eigenvectors for a set of training points $\vec{\alpha}_i$. Any target Hamiltonian ($H(\vec{\alpha}_{\circledcirc})$) is then projected onto this subspace, and the approximate ground state is obtained by diagonalizing a generalized eigenvalue problem. By choosing an optimal set of training points, the ground-state eigenvalue and eigenvector from this subspace rapidly converge to the full-space solution as the number of training points is increased~\cite{sarkar2020,sarkar2022}. In addition, since the target Hamiltonian $H(\vec{\alpha}_{\circledcirc})$ is expressed by a finite number of terms that depend linearly on the \LEC[s], one can project each term $h_i$ onto the subspace such that the resulting matrices can be stored and used to quickly construct the projection of any target Hamiltonian $H(\vec{\alpha}_{\circledcirc})$ onto the subspace of training vectors. 

The proof of convergence of \EC{} emulators as outlined in Ref.~\cite{sarkar2020} has not been generalized to the case of non-Hermitian Hamiltonians as encountered in the \CC{} method. However, the studies of Refs.~\cite{ekstrom2019, Hu:2021trw} showed a rapid convergence to the full-space solution also in this case. 
In this work we employ \NCSM{}-based \EC{} emulators from \textcite{Hu:2021trw} for \nuc{2,3}{H} and \nuc{4}{He} observables. In addition, we construct a new emulator for \nuc{6}{Li} using \textsc{JupiterNCSM}~\cite{Djarv:2021pjc, djarv:jupiterNCSM} to perform model checking. 
For $^{16}$O and nuclear matter we use the \CC{} method~\cite{coester1958,coester1960,cizek1966,kuemmel1978,bishop1991,zeng1998,mihaila2000b,dean2004,bartlett2007,shavittbartlett2009,hagen2010b,binder2013,hagen2014} as the many-body solver and employ the corresponding \SPCC{} method~\cite{ekstrom2019} to construct emulators. For \nuc{16}{O} observables that we use in our statistical analysis we adopt the \SPCC{}-emulators based on the \CC{} method with singles-doubles and leading-order triples excitations (CCSDT-3) as described in Ref.~\cite{Hu:2021trw}. The novel small-batch voting algorithm, that will be presented below, is exclusively used for the nuclear matter \SPCC{} emulators.

\subsection{Subspace-projected coupled-cluster%
\label{sec:SPCC}}
The essence of the \CC{} method is its similarity transformed Hamiltonian:
\begin{eqnarray}
  \label{eq_H}
  \overline{H}(\vec{\alpha}) = e^{-T(\vec{\alpha})} H(\vec{\alpha}) e^{T(\vec{\alpha})},
\end{eqnarray}
where $T(\vec{\alpha})$ is the cluster operator that induces all possible particle-hole excitations. For the nuclear matter emulators in this work we adopt the \CC{} method with doubles (CCD) approximation, so that the cluster operator is truncated at two-particle-two-hole excitations, i.e. $T(\vec{\alpha})=T_2(\vec{\alpha})$. We note that there are no one-particle-one-hole excitations in infinite nuclear matter, i.e., $T_1(\vec{\alpha}) = 0$,  due to momentum conservation. Note that the transformation in Eq.(\ref{eq_H}) is non-unitary and the resulting similarity transformed Hamiltonian $\overline{H}(\vec{\alpha})$ is non-Hermitian. The direct consequence is that the \CC{} method is non-variational, but it follows a bi-variational principle~\cite{arponen1983} by parametrizing the left and right \CC{} ground states as: 
\begin{eqnarray}
  \label{}
  \langle \widetilde{\Psi}|   =\langle \Phi_{0}|(1+\Lambda(\vec{\alpha}))e^{-T(\vec{\alpha})}, \ | \Psi \rangle = e^{T(\vec{\alpha})}  | \Phi_{0} \rangle ,
\end{eqnarray}
where $\Lambda(\vec{\alpha}) = \Lambda_2(\vec{\alpha})$ is a two-hole-two-particle de-excitation operator. We note here that the parametrization of the left state is the first-order approximation to Arponen's extended \CC{} method~\cite{arponen1987} where the left state is parametrized in a more symmetric way by writing $\langle \widetilde{\Psi}|   =\langle \Phi_{0}| e^{\Lambda(\vec{\alpha})} e^{-T(\vec{\alpha})}$. In this work we determined $\Lambda(\vec{\alpha})$ amplitudes by solving the eigenvalue problem
for the left ground state \cite{shavittbartlett2009}.
%
The bi-orthonormality is assured by $\langle \widetilde{\Psi}|\Psi \rangle =1$ when both left and right ground-states are acquired by the same $\vec{\alpha}$. The reference state $| \Phi_{0} \rangle $ is chosen to be the closed-shell configuration on a discrete lattice in momentum space with periodic boundary conditions. The model-space for which we solve the CCD equations has $(2n_{\rm{max}}+1)^3$ momentum points and we set $n_{\rm{max}}=4$, which is sufficiently large to obtain converged results~\cite{hagen2013b}.
To minimize finite-size effects we use 132 nucleons for \SNM{} and 66 neutrons for \PNM{}, respectively~\cite{gandolfi2009,hagen2013b}.

In order to construct the subspace projected target Hamiltonian $H(\vec{\alpha}_{\circledcirc})$ we solve for the left and right \CC{} ground-states for a set of $N_{\rm{sub}}$ training Hamiltonians $H(\vec{\alpha}_{1}),\cdots, H(\vec{\alpha}_{N_{\rm{sub}}})$, and subsequently project $H(\vec{\alpha}_{\circledcirc})$ and the identity matrix onto this subspace giving,
\begin{eqnarray}
  \label{eq_ME}
  \langle \widetilde{\Psi}'|H(\vec{\alpha}_{\circledcirc}) | \Psi \rangle &&=\langle \Phi_{0}|(1+\Lambda')e^{X}\overline{H}(\vec{\alpha}_{\circledcirc})| \Phi_{0} \rangle, \\
   \label{eq_norm}
    \langle \widetilde{\Psi}'| \Psi \rangle &&=\langle \Phi_{0}|(1+\Lambda')e^{X} | \Phi_{0} \rangle,
\end{eqnarray}
where $e^X = e^{-T'+T}$ and we indicate quantities related to different training points, $\vec\alpha$ and $\vec\alpha'$, by unprimed and primed symbols. With Eqs.~(\ref{eq_ME}) and (\ref{eq_norm}) one can easily acquire the ground-state energy for the nuclear matter system by solving a $N_{\rm{sub}} \times N_{\rm{sub}}$ generalized eigenvalue problem. Note that the $N_{\rm{sub}}$ subspace vectors should not be linearly dependent as it would induce numerical instability when solving the generalized eigenvalue problem.

\begin{figure}[htbp]
  \includegraphics[width=1\columnwidth]{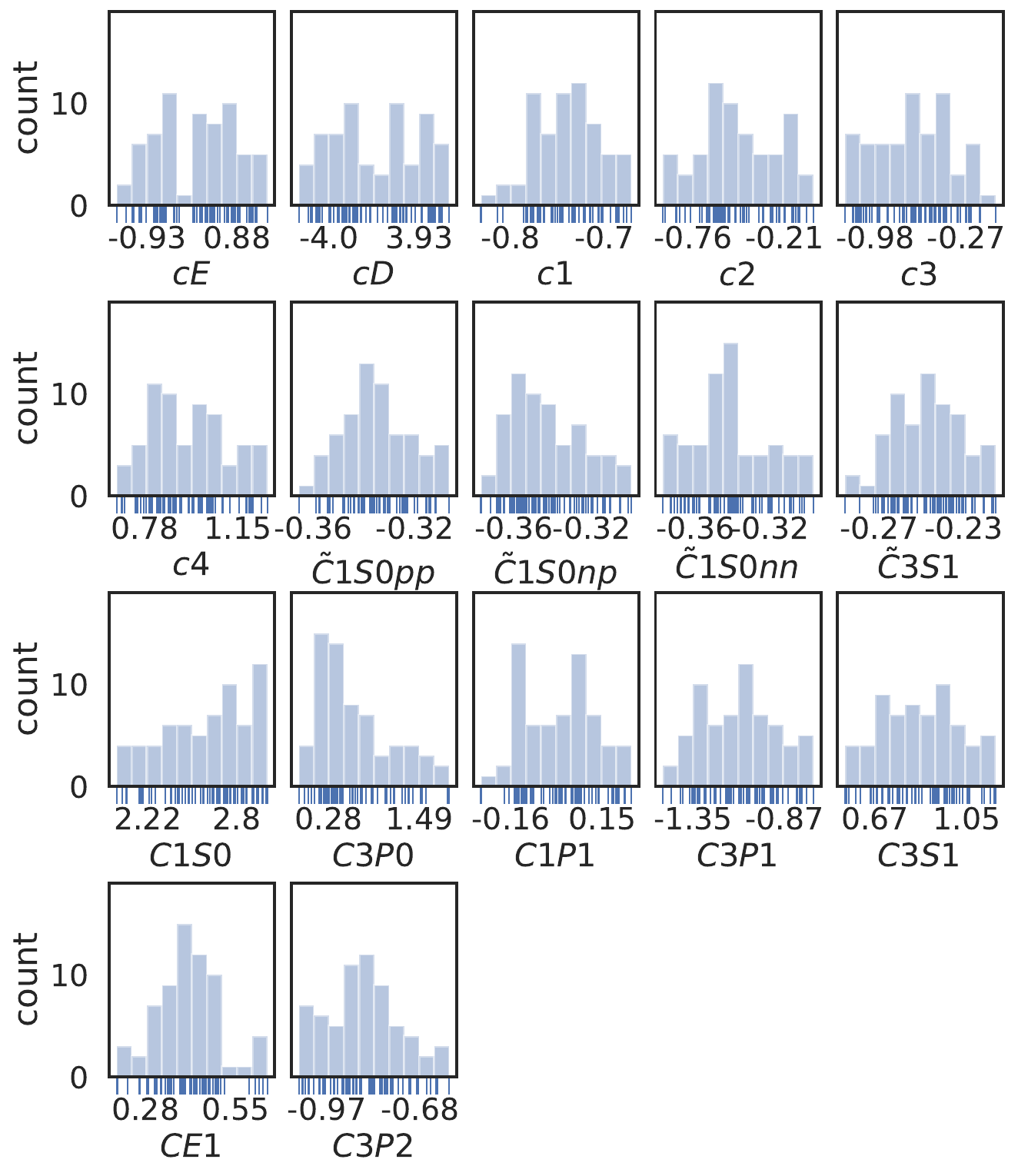}
  \caption{(Color online) Projections of the 64 training points in each \LEC{} dimension (17 different \LEC[s]). Samples are shown both as a histogram and a rug plot (vertical bars under the $x$-axis). The minimum and maximum value for each \LEC{} is printed under each subplot.
  \label{fig:subspace_vector}}
\end{figure}

Another important aspect of the \SPCC{} method is the selection of an appropriate set of training points to construct the subspace. To ensure that the selected training points will lead to accurate emulators in the relevant parameter domain, we first apply history matching to restrict the LEC ranges. Furthermore, to maximize the worth of full computations we select the non-implausible samples with highest likelihood in the final Bayesian analysis as training vectors. More details about the history matching and Bayesian analysis can be found in Sec.~\ref{section:HM} and Sec.~\ref{section:Bayesian}. The training points used in this work are shown in Fig. \ref{fig:subspace_vector} and, as can be seen, they cover a very broad \LEC{} range. 

\begin{figure}[htbp]
  \includegraphics[width=0.7\columnwidth]{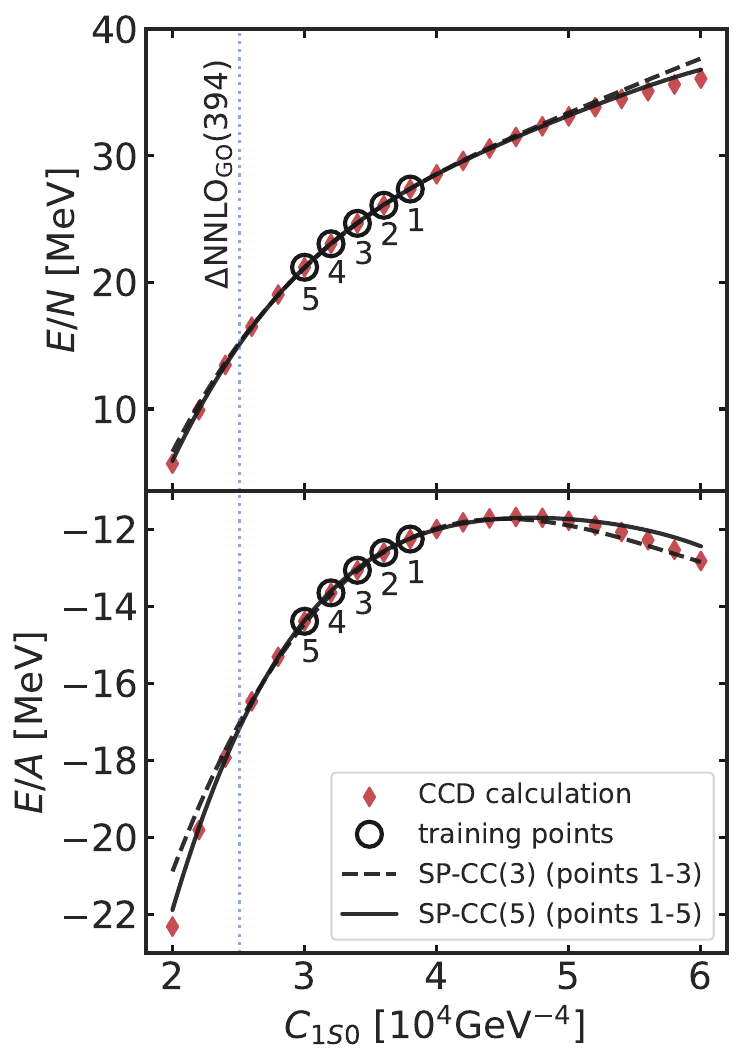}
  \caption{(Color online) Demonstration of \SPCC{} predictions of \PNM{} and \SNM{} at $\rho=0.16 \ \rm{fm^{-3}}$, using three or five training points (circle markers), for different values of the low-energy constant $C_{1S0}$. The other \LEC[s] are fixed at their values in $\Delta$NNLO$_{\rm{GO}}(394)$~\cite{jiang2020}. The red diamonds correspond to exact CCD calculations. The $C_{1S0}$ value of $\Delta$NNLO$_{\rm{GO}}(394)$ is indicated with a dashed vertical line. This one-dimensional emulator demonstration is obtained for a model with $N=14$ ($A=28$) for \PNM{} (\SNM{}).
  \label{fig:emulator_vs_ccd_one_parameter}}
\end{figure}

\subsection{Emulators for a single LEC}
We first consider an illustrative example with simpler nuclear matter emulators. For this purpose we use an interaction model with a single free parameter and perform nuclear-matter modeling with a smaller number of particles. Specifically, we employ the $\Delta$NNLO$_{\rm{GO}}(394)$ interaction~\cite{jiang2020} and allow the single $C_{1S0}$ \LEC{} to vary and we use 14 (28) neutrons (nucleons) for \PNM{} and \SNM{}, respectively.
Fig. \ref{fig:emulator_vs_ccd_one_parameter} shows the calculated energy per neutron ($E/N$) and energy per nucleon ($E/A$) for \PNM{} and \SNM{}, respectively. The \SPCC{} predictions using three or five subspace vectors are compared with full space CCD results for a wide range of the low-energy constant $C_{1S0}$ (the remaining \LEC[s] are kept fixed). As we can see, using $N_{\rm{sub}}=5$ training points chosen in a small region, the \SPCC{} method already accurately reproduces the full space CCD calculations over a large range for the $C_{1S0}$ \LEC{}. As expected, if we reduce the number of training points to $N_{\rm{sub}}=3$, the \SPCC{} predictions of \SNM{} start to deviate more from the exact solutions in the case of large exptrapolations. However, the predictions for \PNM{} remain accurate over the whole range considered. The choice of training points in Fig. \ref{fig:emulator_vs_ccd_one_parameter} is just used for illustration. When constructing actual emulators the training points cover a larger range of the given parameter space to achieve better performance.

\subsection{Small-batch voting} \label{section:SBV}
When building \SPCC{} emulators for systems with 132 nucleons for  \SNM{} one suffers from a persistent spurious state problem. We find that there can be several eigenstates of the $N_{\rm{sub}} \times N_{\rm{sub}}$ matrix that have lower eigenvalues than the corresponding full-space CCD result.
The exact reason for the appearance of these states is not yet fully understood, but it could be a consequence of several factors: (i) the \SPCC{} Hamiltonian is by construction non-Hermitian and the variational theorem does not apply; (ii) for increasing number of nucleons (132 nucleons and 66 neutrons in our case) the level density increases which more easily leads to the occurence of these states, and (iii) increasing correlation energies associated with less perturbative interactions might produce more spurious states.

As we consider these spurious states to be unphysical we seek a method to identify them \emph{a priori} in order to remove them from the spectrum.
Recall that \CC{} theory fulfills a bivariational theorem and the physical solution is a stationary point with respect to variations of the \CC{} amplitudes.  Whether the bivariational property of \CC{} theory also holds for the SPCC remains to be shown, but it is reasonable to assume that it holds as long as the subspace is sufficiently large. In this section we will show how we can use the bivariational property to efficiently identify the physical solution within the \SPCC{} spectrum using a method we call small-batch voting.

\begin{figure}[htbp]
  \includegraphics[width=0.95\columnwidth] {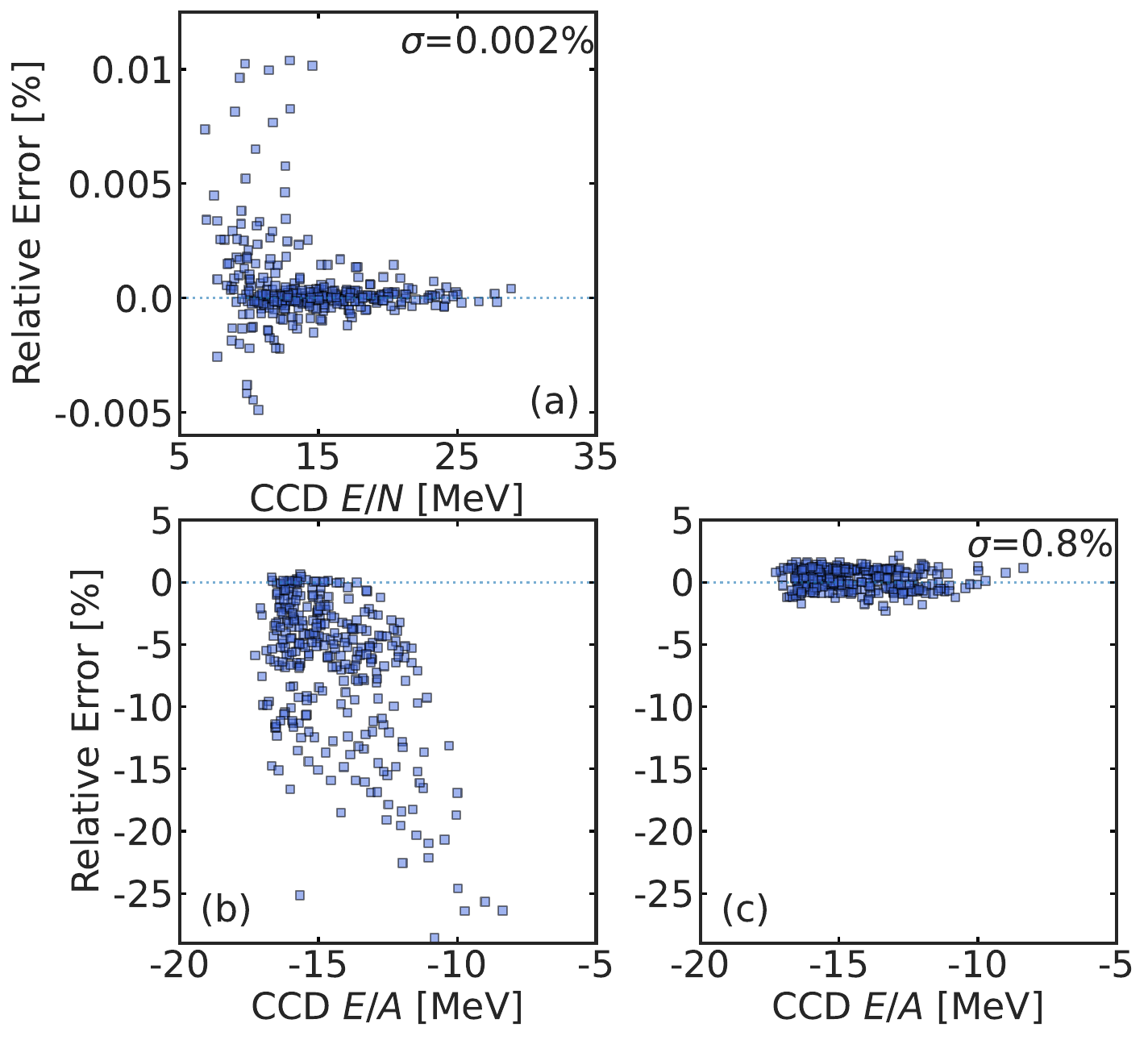}
  \caption{(Color online) 
  Relative errors between \SPCC{} predictions and exact CCD calculations for \PNM{} (top) and \SNM{} (bottom). Panels (a) and (b) show validation results without small-batch voting. For \SNM{} we use small-batch voting for the final emulator. The significantly improved validation results are shown in panel (c).
  \label{fig:emulator_vs_ccd}}
\end{figure}

Fig. \ref{fig:emulator_vs_ccd}(a)(b) illustrates the relative errors $(E_{\rm{SPCC}}-E_{\rm{CCD}})/|E_{\rm{CCD}}|$ between emulator predictions and exact CCD calculations for \PNM{} (\SNM{}) modeled with 66 (132) neutrons (nucleons). The emulator predictions are chosen as the SPCC solution with the lowest (real) energy. The validation points correspond to 50 random parametrizations of the $\Delta$NNLO(394) interaction with the energy per particle computed for five densities: $\rho \in \{0.12, 0.14, 0.16, 0.18, 0.20\}$~fm$^{-3}$. Thus, there are 250 points in total. In these calculations, $N_{\rm{sub}}=64$ subspace vectors are used to construct the \SPCC{} emulators. Note that the validation interactions are selected randomly within a constrained parameter domain (resulting from history matching). For \SNM{}, most errors are negative which indicates that the emulator predictions correspond, in fact, to spurious states that give lower energies than the corresponding full-space \CC{} calculations. Based on this, it seems the occurrence of spurious states seriously hampers the predictive power of the \SPCC{} method for \SNM{}. The fact that the \SPCC{} method works better for \PNM{} is probably due to the smaller correlation energies for this system, i.e., it is more perturbative.

However, when comparing the full eigenspectrum of the subspace problem to the full-space \CC{} result, we find that the physical ground-state indeed is contained therein. The challenge then is to efficiently identify this state without actually doing any full-space \CC{} calculations.
Assuming that  the subspace is large enough to accurately describe the exact \CC{} ground-state solution, the eigenvector should also fulfill the bivariational principle. This means we can write the exact solution as,
\begin{equation}
  \vert \Psi (\vec{\alpha}_{\circledcirc}) \rangle =e^{T(\vec{\alpha}_{\circledcirc})} \vert \Phi_0\rangle  \approx \sum_{i=1}^{N_{\rm{sub}}} c_i^{\star} e^{T(\vec{\alpha}_i)} \vert \Phi_0\rangle, 
\end{equation} 
here $\vec{c^{\star}}$ is the \SPCC{} eigenvector that corresponds to the physical ground-state (denoted by $\star$). The physical solution converges rapidly with increasing number of training points. By utilizing the bivariational property, it is then reasonable to assume that the physical state should remain stable (displaying small variations in the corresponding eigenenergy) when removing a small portion $\Delta N$ of the subspace, while the spurious states and their energies should change significantly. Here we introduce batches, $N_{\rm{batch}} =  N_{\rm{sub}} - \Delta N$, that should be large enough to still provide an accurate representation of the physical ground state. 

Based on this argument we develop a new algorithm called small-batch voting to efficiently identify the physical ground state and counter the spurious-state problem. The procedure of small-batch voting can be summarized as follows:
\begin{enumerate}
  \item Solve the target Hamiltonian $H(\vec{\alpha}_{\circledcirc})$ using the \SPCC{} method in a subspace with relatively large $N_{\rm{sub}}$ to ensure that the physical ground-state is well established in the spectrum. The eigenvalues of $H(\vec{\alpha}_{\circledcirc})$ are stored as $E_{i=1,...N_{\rm sub} }$
  \item Construct $k$ different small batches by randomly picking subsets of $N_{\rm{batch}}  (< N_{\rm{sub}}$) vectors from the original subspace. 
  \item For each batch, solve the generalized $N_{\rm{batch}}   \times N_{\rm{batch}}  $ eigenvalue problem. Compare the eigenvalues $e_{r=1,...N_{\rm batch} }$ with the eigenvalues of the original subspace $E_{i=1,...N_{\rm sub} }$. If the same eigenvalue occurs in both spectra (within a specified relative tolerance $|(E_i-e_{r})/E_i|<\varepsilon$), the corresponding eigenvector of the original $N_{\rm{sub}} \times N_{\rm{sub}}$ subspace gets one vote, $v_{i} = v_{i}+1$.
  \item Repeat step 3 for all $k$ small batches.
  \item The eigenvector with the highest number of votes $v^* = \max (v_{i})$ is assumed to correspond to the physical ground state (with the lowest energy as a deciding vote if there is a draw) and its eigenvalue $E^*$ is used as the emulator prediction.    
  \end{enumerate}
In this work, we set $N_{\rm sub}=64$, $k=100$, $N_{\rm{batch}}=30$, and the relative tolerance $\varepsilon = 0.02$. We checked that $N_{\rm{batch}}=30$ is sufficiently large to reproduce the fullspace \CC{} solution to within $1\%$ when we know exactly which is the physical ground-state in the spectrum.
To summarize, the essential idea of the algorithm is that by varying the composition of the subspace, the eigenvalues of the spurious states will be shifted dramatically since they are not stationary solutions, while the physical ground-state remains relatively unchanged.

Fig.~\ref{fig:emulator_vs_ccd} summarizes our results with and without small-batch voting. 
Panels (a) and (b) show the relative error between the lowest \SPCC{} energies (without small-batch voting) and the corresponding exact CCD results, while panel (c) show the results of \SNM{} with small-batch voting. 
For the latter we apply a 1\% mean shift up in energy since the voting procedure favors the lowest state that is found within the 2\% tolerance window.
The comparison demonstrates that the small-batch voting algorithm successfully removes most of the spurious states and that the emulator predictions are much improved. Note that for \PNM{} the \SPCC{} predictions are already extremely accurate thus we do not apply small batch voting for the \PNM{} emulator.
The standard deviation of the relative error is $\sigma=0.002\%$ ($0.8\%$) for \PNM{} (\SNM{}). We note that there are still a few spurious states that remain after applying the small-batch voting algorithm for the more complex non-perturbative \SNM{} case. The total computational cost of the SPCC emulators for \SNM{} with small-batch voting is six orders of magnitude smaller than the corresponding full space \CC{} calculations. For \PNM{} emulators, where we don't need small-batch voting, we have a computational speedup of more than eight orders of magnitude.

\subsection{Nuclear matter saturation observables%
\label{sec:NMsaturation}}
At this point we are able to construct emulators for \PNM{} and \SNM{} and predict the energy per particle at a given density using the \SPCC{} method with small-batch voting. In order to study nuclear matter saturation properties, e.g., the saturation density $\rho_0$, the saturation energy $E_{0}/A$, the symmetry energy $S$, the symmetry energy slope parameter $L$ and the incompressibility $K$, one needs to acquire the \EOS{} for both pure neutron and symmetric nuclear matter around the saturation point. Ideally one would like to include the density $\rho$ parameter in the eigenvector continuation scheme and build an emulator that works for different \LEC[s] and at arbitrary densities. However, changing the density leads to different discretizations of the momentum space lattice and one would therefore need to work out matrix elements connecting different reference states and lattices. 

Fortunately, we are not completely ignorant about the properties of the \EOS{} of nuclear matter since the energy per particle should be represented by continuous smooth functions of $\rho$. This smoothness implies that we do not need many density points to obtain sufficient information about the \EOS{}. In this work, we construct SPCC emulators for both \PNM{} and \SNM{} at five different densities:  $\rho = 0.12,\,014,\,0.16,\,0.18,\,0.20$~fm$^{-3}$. We choose to study this density region simply because the empirical saturation density is around $0.16 \ \rm{fm}^{-3}$ \cite{hebeler2011,bender2003}. The nuclear matter \EOS{} is then interpolated within this range by using \GP[s]~\cite{rasmussen2006} as the interpolation method. We choose the radial basis function (RBF) as the correlation function to ensure the smoothness of the \EOS{}. The hyperparameter (correlation length $l$) of the \GP{} is learned from a validation data set which contains 50 interaction samples that are generated by the same history matching process mentioned in Sec.~\ref{section:SBV}. The \PNM{} and \SNM{} correlation lengths optimized from the validation set are $0.297\ \rm{fm}^{-3}$ and $0.259 \ \rm{fm}^{-3}$, respectively. In the end, we take the more conservative value $l=0.25\ \rm{fm}^{-3}$ for both \PNM{} and \SNM{} so that we do not overestimate the correlation length. 

\begin{figure}
  \includegraphics[width=0.9\columnwidth] {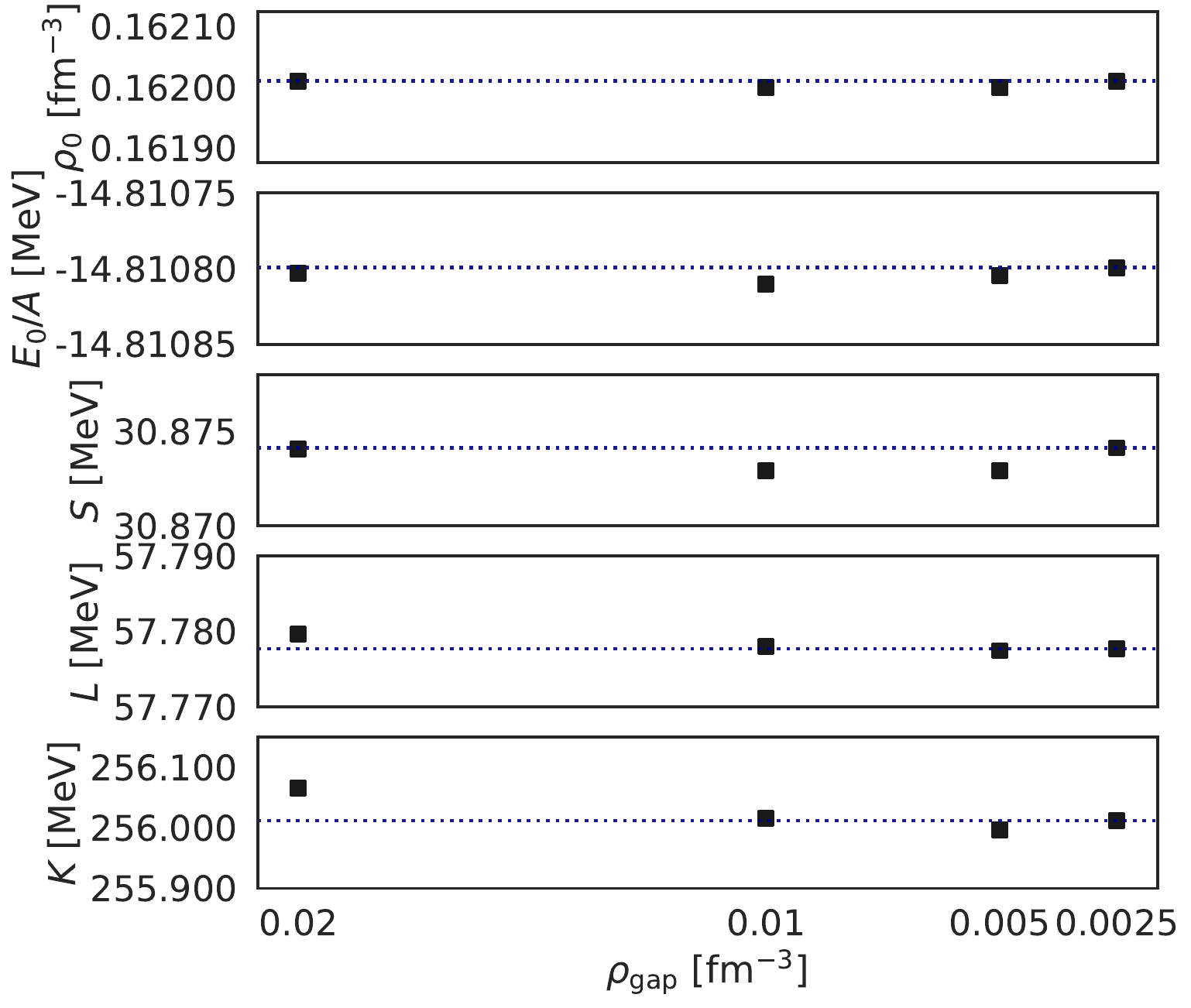}
  \caption{(Color online) The convergence of saturation properties extracted by \GP{} interpolation with different density spacings. The hyperparameter of the RBF kernel used in the \GP{} is $l=0.25 \ \rm{fm^{-3}}$.  In this convergence study, the energy per particle at discrete density points is obtained with the \CC{} method for a single interaction. 
  \label{fig:GP}}
\end{figure}

The major advantage of using \GP[s] is that they are infinitely differentiable under the RBF kernel and that derivative properties such as $L$ and $K$ can easily be obtained. For any given interaction with its saturation point $\rho_0 \in [0.12, 0.20]$~fm$^{-3}$ we can therefore extract all saturation properties from the corresponding \GP[s] and its (first and second) derivatives.

Fig.~\ref{fig:GP} shows the performance of \GP{} interpolation with different density spacing $\rho_{\rm gap}$ from $0.02$ to $0.0025$ $\rm{fm^{-3}}$. It can be seen that $L$ and $K$, which correspond to first and second derivatives of the \EOS{}, have larger deviation when the density spacing is increased. In principle, we expect better accuracy with smaller density spacing. However, the values of saturation properties at $\rho_{\rm gap} = 0.02 \ \rm{fm^{-3}}$ differ by less than $0.1\%$ compared with $\rho_{\rm gap} = 0.0025 \ \rm{fm^{-3}}$. This difference is rather small compared to other sources of uncertainty. In practice we therefore use $\rho_{\rm gap}=0.02 \ \rm{fm^{-3}}$ and ignore the GP interpolation error.

\subsection{History matching%
  \label{section:HM}}
In this work we use an iterative history matching approach~\cite{vernon2010,vernon2014,vernon2018, Hu:2021trw} with selected experimental few-nucleon data to study and reduce the huge parameter space of our \chiEFT{} interaction model.
For each wave of history matching we need to establish a quantitative criterion that determines if a parametrization $\vec{\alpha}$ yields acceptable (or at least not implausible) model predictions when confronted with the selected set of observations $\mathcal{Z}$. We first introduce the individual implausibility measure
\begin{equation}\label{individual_implausibility_measure}
	I^2_i (\vec{\alpha}) = \frac{\left| M_i(\vec{\alpha}) - z_i  \right|^2}{\mathrm{Var} \left( M_i(\vec{\alpha}) - z_i  \right)  },
\end{equation}
which includes the squared difference between the model prediction $M_i(\vec{\alpha})$ and the observation $z_i$ for observable $i$ from the target set $\mathcal{Z}$. The total variance in the denominator of Eq.~(\ref{individual_implausibility_measure}) is here constructed under the assumption of independent errors. It is therefore a sum of variance terms that in our case include experimental, model, method, and emulator errors. Unless differently specified we use the maximum of the individual measures to define the implausibility constraint
\begin{equation}\label{maximum_implausibility}
I_M (\vec{\alpha}) \equiv \mathop{\mathrm{max}}\limits_{z_i \in \mathcal{Z}} I_i (\vec{\alpha}) \leq c_I,
\end{equation}
where the default choice is $c_I \equiv 3.0$ inspired by Pukelheim's three-sigma rule~\cite{Pukelsheim:1994}.

\begin{figure*}[!htb]
\includegraphics[width=0.80\linewidth]{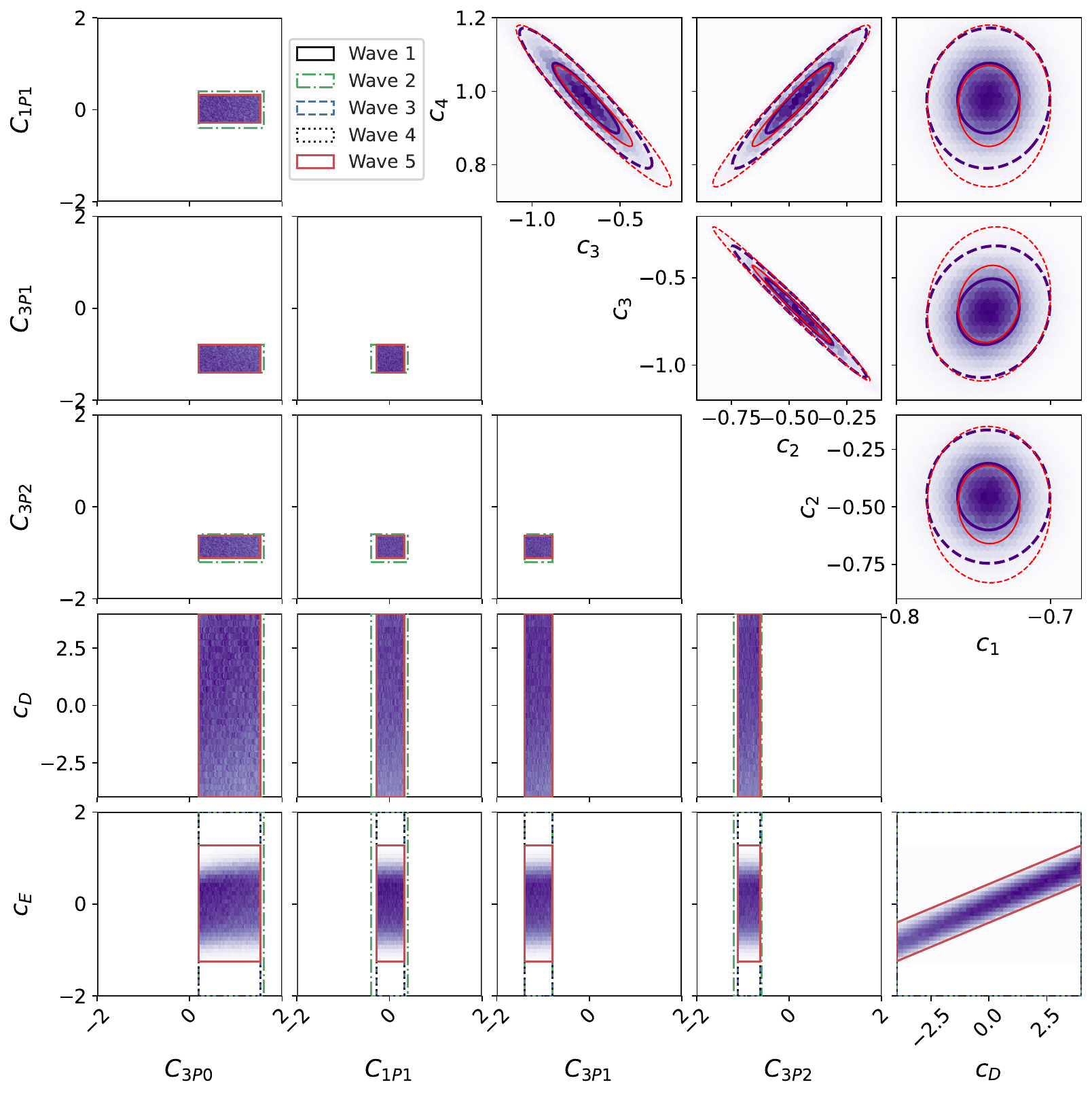}
\caption{(Color online) Non-implausible parameter domains for ($C_{1P1}$, $C_{3P0}$, $C_{3P1}$, $C_{3P2}$, $c_D$, $c_E$ and $c_{1,2,3,4}$) at the ends of the five history matching waves. The initial parameter domain is represented by the axes limits for all panels (except $c_{1,2,3,4}$). The volume of the non-implausible domain is iteratively reduced in waves 2, 3, 4, and 5 (shown by green/dash-dotted, blue/dashed, black/dotted, red/solid rectangles, respectively). The non-implausible samples in the final wave are shown as two-dimensional histograms (purple). Note that the sampled volume for $c_{1,2,3,4}$ (illustrated by red/solid and dashed contour lines denoting $68\%$ and $90\%$ credible regions, respectively) remain the same in all waves. In practice, for these \LEC[s], we use a four-dimensional hypercube mapped onto the multivariate Gaussian probability density function resulting from a Roy-Steiner analysis of pion-nucleon scattering data~\cite{siemens2017}. The contour lines (purple/solid and dashed) for the non-implausible samples identified in the final wave are shown for comparison.
\label{fig:HM_waves_one}
}
\end{figure*}

\begin{figure*}[!htb]
\includegraphics[width=0.80\linewidth]{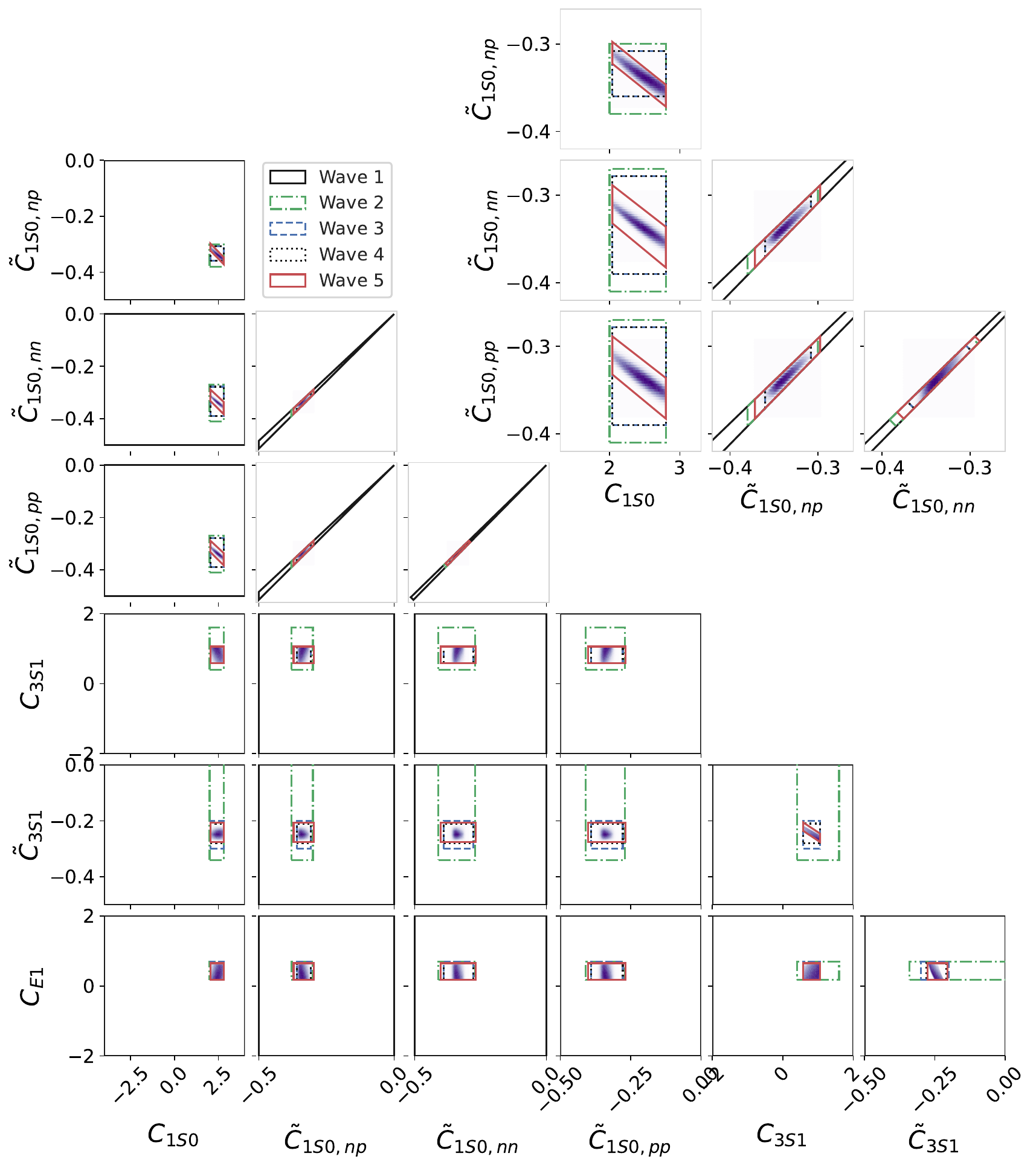}
  \caption{(Color online)  Non-implausible parameter domains for ($\tilde{C}_{1S0,np}$, $\tilde{C}_{1S0,nn}$, $\tilde{C}_{1S0,pp}$, $\tilde{C}_{3S1}$, $C_{1S0}$, $C_{3S1}$, $C_{E1}$) at the ends of the five history matching waves. Note that the input volume for the isospin symmetry breaking \LEC[s] ($\tilde{C}_{1S0,np}$, $\tilde{C}_{1S0,nn}$, $\tilde{C}_{1S0,pp}$) are strongly correlated. The initial parameter domain is displayed by the black/solid quadrilaterals. The volume of the non-implausible domain is iteratively reduced in waves 2, 3, 4, and 5 as shown by the green/dash-dotted, blue/dashed, black/dotted, red/solid quadrilaterals. The non-implausible samples in the final wave are illustrated as two-dimensional histograms (purple). Enlargements of the most relevant region of the \LEC{} pairs of ($\tilde{C}_{3S1}$, $\tilde{C}_{1S0,np}$, $\tilde{C}_{1S0,nn}$, $\tilde{C}_{1S0,pp}$) are shown in the top right panels. 
\label{fig:HM_waves_two}
}
\end{figure*}

History matching proceeds by reducing the parameter space iteratively. In each wave one removes regions that are deemed implausible by failing the constraint~in Eq.~\eqref{maximum_implausibility} .
\begin{figure}[!htb]
\includegraphics[width=1\linewidth]{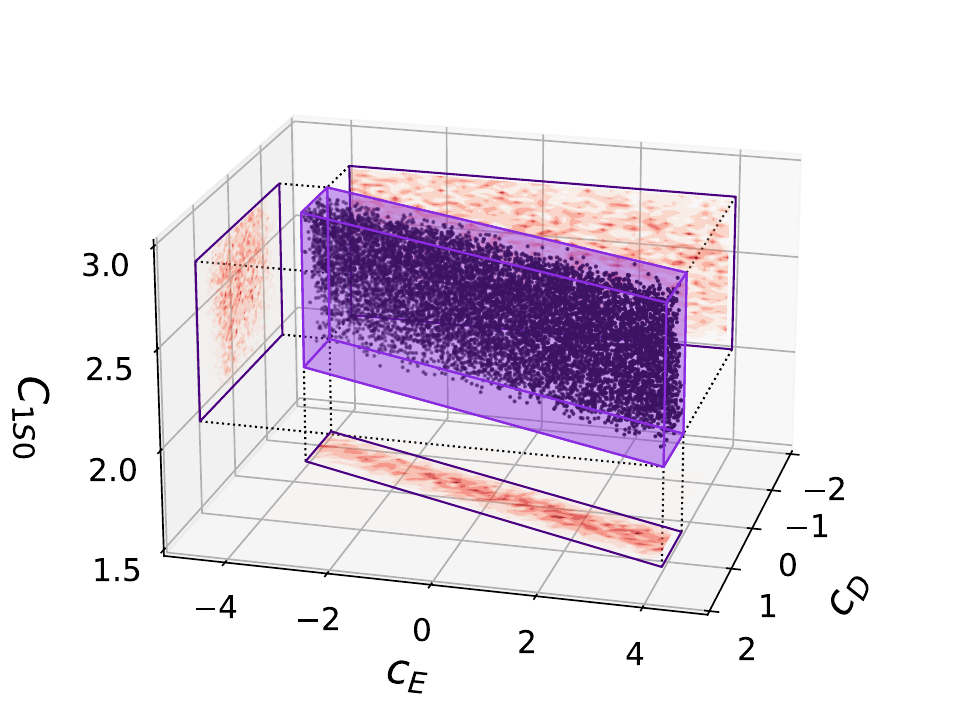}
\caption{(Color online) Visualization of the history matching procedure in a trivariate $(C_{1S0}, c_D, c_E)$-subspace. The \LEC[s] \cD{} and \cE{} are dimensionless while $C_{1S0}$ is in units of $10^4$ {GeV}$^{-4}$. The non-implausible interaction samples are shown as dark dots. These dots are projected on different \LEC{} surfaces and the outlines of the bounding regions are represented by parallelograms. The purple box outlines the final non-implausible \LEC{} domain.
    \label{fig:HM}
}
\end{figure}
A visualization of this process is shown in Fig.~\ref{fig:HM}. We first use a space-filling Latin hypercube design~\cite{santner2003design} to generate well-spaced interaction samples in the input parameter domain. Then we use fast modeling or emulation to compute the implausibility measures and apply the maximum implausibility constraint. The remaining non-implausible interaction samples are kept and define the non-implausible region for the next wave.  Note that we are using parallelograms to define the bivariate surfaces of the non-implausible volume which allows to incorporate parameter correlations.
In this work the iterative history matching is carried out in five waves as shown in Table~\ref{tab:HM_waves}. Initial waves comprise selected groups of observables and subsets of active input parameters. This enables reaching sufficiently high resolution in the space-filling design of interaction samples.
\begin{table*}[!htb]
 \caption[HM waves]
 {Properties and summary statistics of the five waves of history matching performed in this work. See the text for details on the few-nucleon observables that are included in the target sets. The number of active inputs correspond to the dimensionality of the \LEC{} subspace that is being explored in each wave. The second to last column indicates the fraction of input samples that passed the implausibility test, while the last column shows how large proportion of the initial volume that remains.
 \label{tab:HM_waves}%
}
    \begin{tabular}{ccccccc}
    \hline
    & \multicolumn{2}{c}{Target set $\mathcal{Z}$} & Active & Input & Non-implausible & Proportion space\\
   Wave & outputs & systems & inputs & samples & fraction & non-implausible \\
    \hline
    1 & $6 \times 6$ & $np$ scattering & 5--7 & {$ 10^6$--$2.7 \times 10^8$} &  {$10^{-1}$--$10^{-4}$} &$1.5 \times 10^{-6}$\\
    2 & $6 \times 6$ & $np$ scattering &  5--7 & {$10^6$--$2.7 \times 10^8$} & {$10^{-1}$--$10^{-4}$} &$3.7 \times 10^{-8}$\\ 
    3 & 3 & $A=2$ & 7 & {$2.7 \times 10^8$} & {$7 \times 10^{-3}$} &$2.4 \times 10^{-8}$ \\ 
    4 & 6 & $A=2$--$4$ & 13 & {$10^8$} & {$1.3 \times 10^{-4}$}  &$1.0 \times 10^{-9}$\\ 
    5 & 6 & $A=2$--$4$ & 17 & {$ 10^9$}& {$1.7 \times 10^{-3}$}  &same\\ 
   \hline
    \end{tabular}
\end{table*}
Figs.~\ref{fig:HM_waves_one} and~\ref{fig:HM_waves_two} show the non-implausible volume for each wave including the final one. In waves 1 and 2 we constrain all relevant \LEC[s] (except \cD{} and \cE{}) grouped by partial waves ($^1S_0$, $^3S_1$, $^1P_1$, $^3P_0$, $^3P_1$, $^3P_2$) using neutron-proton scattering phase shifts at six energies ($T_{\rm lab}= 1,5,25,50,100,200$). In wave 3 we include only the deuteron ground-state energy, point-proton radius and quadrupole moment as target observables and consider $\tilde{C}_{3S1}$, $C_{3S1}$, $C_{E1}$ and $c_{1,2,3,4}$ as active parameters. In wave 4, we add the \nuc{3}{H} binding energy and the \nuc{4}{He} binding energy and point-proton radius to the set of target observables. Here we consider also
the three-nucleon force parameters $c_D$, $c_E$ along with the other \LEC[s]. In this wave, however, we fixed the four $P$-wave \LEC[s] ($C_{1P1}$, $C_{3P0}$, $C_{3P1}$, $C_{3P2}$) to the values from the $\Delta$NNLO$_{\rm{GO}}(394)$ interaction~\cite{jiang2020}, since the selected target observables are not very sensitive to these parameters. We added an additional method uncertainty to the denominator of the implausibility metric~\eqref{individual_implausibility_measure} to capture the reduced precision of the model with fixed $P$-wave parameters. Following a sensitivity study we set the standard deviation of this additional error to 100 and 400~keV for the \nuc{3}{H} and \nuc{4}{He} binding energy, respectively, and to 0.03~fm$^2$ for the squared point-proton radius of \nuc{4}{He}.

Finally in wave 5, all 17 model parameters are active and the non-implausible domain is explored by $1\times 10^9$ space-filling design samples. In the end we find that $1.7\times 10^6$ of them pass the implausibility constraint with the same set of few-nucleon target observables as in wave 4. It is worth to mention that the order at which observables are considered in history matching waves is irrelevant as long as one uses the maximum implausibility as the constraint. With the maximum implausibility measure, the final non-implausible region is the intersection of constrained parameter regions from different waves and is therefore unrelated to wave ordering.
\subsection{Bayesian machine learning error model} \label{section:error}
In order to connect the emulator predictions with actual nuclear matter properties one needs to incorporate errors from different sources.
A statistical model for the density-dependent energy per particle, that accounts for the most relevant sources of uncertainty, can be written as
\begin{equation}
y(\rho) = y_k(\rho) + \varepsilon_k(\rho) + \varepsilon_{\rm{method}}(\rho)+ \varepsilon_{\rm{emu}}(\rho),
\label{eq:predicted_observables}
\end{equation}
where $y_k(\rho)$ is the nuclear matter emulator prediction using our \EFT{} model truncated at order $k$ at given density $\rho$, while the stochastic terms $\varepsilon_k(\rho)$, $\varepsilon_{\rm{method}}(\rho)$, and $\varepsilon_{\rm{emu}}(\rho)$ correspond to the \EFT{} truncation error (model discrepancy), the method error, and the emulator error, respectively. To quantify the distributions of these stochastic variables we apply and extend a Bayesian machine learning error model that was originally proposed by~\textcite{drischler2020a}. Within this error model we construct multitask \GP[s]~\cite{alvarez2012} to estimate both the variance and the covariance of target errors as a function of density and proton-to-neutron fraction (\SNM{} and \PNM{}) from given prior information.

Let us first consider the \EFT{} truncation error. We follow Refs.~\cite{epelbaum2015,furnstahl2015} and write the \EFT{} expansion for an observable, truncated at order $k$, as
\begin{equation}
 y_k(\rho)=  y_{\rm{ref}}(\rho) \sum^{k}_{n=0} c_n(\rho)Q^n(\rho).
\label{eq:EFT_observable}
\end{equation}
Here $y_{\rm{ref}}$ is a reference scale for the observable $y$, $c_n$ are dimensionless expansion coefficients (with $c_1$ = 0 in Weinberg power counting), while the expansion parameter $Q=k_F/\Lambda_b$ is the ratio of the Fermi momentum $k_F$ and the breakdown momentum $\Lambda_b$. Here we use $\Lambda_b=600$~MeV. Note that for simplicity we fixed $\Lambda_b$ and used the Fermi momentum as the energy scale for the \EOS{}. In future studies one could attempt to infer $\Lambda_b$ or $Q$ directly when making inferences to data that is more sensitive to higher energies. With a simple extension of Eq.~\ref{eq:EFT_observable} to infinite order, the truncation error $\varepsilon_k$ at order $k$ can be expressed as
\begin{equation}
\varepsilon_k(\rho)= y_{\rm{ref}}(\rho) \sum^{\infty}_{n=k+1} c_n(\rho)Q^n(\rho)
\label{eq:EFT_truncation_error}
\end{equation}
We will infer the expansion coefficients $c_n(\rho)$ given \EFT{} convergence assumptions and choose $y_{\rm{ref}}(\rho) = y_\text{LO}(\rho)$. We further assume that $c_i(\rho)$ and $c_j(\rho)$, at different orders $i$ and $j$, should be independent and identically
distributed random functions of natural size. Thus, the error model assumes that they can be described by a single underlying \GP{}
\begin{equation}
c_n(\rho) ~ |~ \bar{c}^2,l \sim GP[0,\bar{c}^2 r(\rho,\rho';l)],
\label{eq:Cn}
\end{equation}
where we use a (Gaussian) radial basis correlation function $r(\rho, \rho'; l)$ with $\bar{c}^2$ and $l$ the \GP{} hyperparameters corresponding to the variance and the correlation length, respectively. Note that in these equations $\rho$ and $l$ are measured in fm$^{-1}$ as we translate from density to the corresponding Fermi momentum for each type of nuclear matter. The mean function of the \GP{} is taken to be zero since the correction at each order can be positive or negative. With Eqs.~\eqref{eq:EFT_truncation_error} and \eqref{eq:Cn} one can easily derive the \EFT{} truncation error as a geometric sum of independent normally distributed variables. Its distribution then follows
\begin{equation}
\varepsilon_k(\rho) ~ |~ \bar{c}^2,l,Q \sim GP[0,\bar{c}^2R_{\varepsilon_k}(\rho,\rho';l)],
\label{eq:truncation_error_1}
\end{equation}
with
\begin{equation}
R_{\varepsilon_k}(\rho,\rho';l) = y_{\rm{ref}}(\rho)y_{\rm{ref}}(\rho')\frac{[Q(\rho)Q(\rho')]^{k+1}}{1-Q(\rho)Q(\rho')}r(\rho,\rho';l).
\label{eq:truncation_error_2}
\end{equation}
We note that $k=3$ for $\Delta$NNLO.

Having defined the \GP{} that describes the truncation error, the hyperparameters $\bar{c}^2$ and $l$ can be inferred from order-by-order \EFT{} predictions and expert elicitation~\cite{santner2003design}.
Using data $\mathcal{D}$, corresponding to order-by-order predictions at several densities $\rho$, and the incorporation of \EFT{} expectations via a prior ${\rm{pr}}(\bar{c}^2,l)$, the posterior for $\bar{c}^2$ and $l$ becomes
\begin{equation}
{\rm{pr}}(\bar{c}^2,l~|~\mathcal{D}) \propto \mathcal{L}(\mathcal{D}~|~\bar{c}^2,l){\rm{pr}}(\bar{c}^2,l).
\label{eq:posterior_hyperparameter}
\end{equation}
Here we use a scaled inverse-chi-squared distribution~\cite{melendez2019} as the prior for $\bar{c}^2$ and a uniform prior for $l$ (with $l\in[0,10]$).
In practice, we first train two \GP[s] separately with order-by-order predictions of the \EOS{} for \PNM{} and \SNM{} (at discrete density points $\rho = 0.06,0.14,... 0.38$~fm$^{-3}$) using \LO{}, \NLO{} and \NNLO{} $\Delta$-full interactions~\cite{Andreas:private} that were optimized using the protocol described in Ref.~\cite{ekstrom2017}.
See the Supplemental Material~\cite{supp2022:PRC} for numerical values of the \LEC[s] that define these convergence-study interactions.
For simplicity we then used the maximum a posteriori (MAP) value as a point estimate for the hyperparameters. The hyperparameters learned in this way from the training data are $\bar{c}_1=0.99$ and $l_1= 0.88~\rm{fm}^{-1}$ for \PNM{} and $\bar{c}_2=1.66$ and $l_2= 0.45~\rm{fm}^{-1}$ for \SNM{}. 

These two \GP[s] (separately trained for \PNM{} and \SNM{}) describe the correlation of truncation errors as a function of density for either system individually. As discussed in Refs.~\cite{drischler2020a,drischler2020b} it is crucial to also account for the cross correlation between \PNM{} and \SNM{} truncation errors. This is important to avoid overestimating the total uncertainty for observables such as the symmetry energy $S$ that corresponds to the difference between $E/N$ and $E/A$. Applying a multitask \GP{} model, the truncation errors of \PNM{} and \SNM{} become jointly distributed
\begin{equation}
\begin{bmatrix}
\varepsilon_{k,\rm{pnm}}\\
\varepsilon_{k,\rm{snm}}
\end{bmatrix} \sim N \left(
\begin{bmatrix}
0\\
0
\end{bmatrix},
\begin{bmatrix}
K_{11}& K_{12}\\
K_{21}& K_{22}
\end{bmatrix}  \right) ,
\label{eq:multitask_GP}
\end{equation}
where $K_{ii}$ is the covariance matrix generated by the kernel function $\bar{c}_i^2R_{\varepsilon_k}(\rho,\rho';l_i)$ of \PNM{} ($i=1$) and \SNM{} ($i=2$) respectively, as described above, while $K_{12} = K_{21}^{T}$ is the cross-covariance that we describe with the kernel function $\rho_{12} \bar{c}_1\bar{c}_2 R_{\varepsilon_k}(\rho,\rho';l_{12})$. Following Ref.~\cite{drischler2020b} we set the cross correlation coefficient $\rho_{12} = \sqrt{2l_1l_2/(l_1^2+l_2^2)}=0.90$ and correlation length $l_{12}=\sqrt{(l_1^2+l_2^2)/2}=0.70$.

In this work we extend the statistical error model by also incorporating the method error and its correlation structure into the analysis.
Specifically we consider our two main sources of method uncertainty, namely the truncation of the cluster operator and the finite-size effect of the cubic momentum lattice. Assuming independence, the total method error can then be written as $\varepsilon_{\rm{method}}=\varepsilon_{\rm{cc}}+\varepsilon_{\rm{fs}}$. We will again use the \GP{} error model such that 
\begin{equation}
\varepsilon_\kappa(\rho) ~ |~ \bar{c}_{\kappa}^2,l_{\kappa}, \sim GP[\mu_{\kappa}(\rho),\bar{c}_{\kappa}^2R_{\kappa}(\rho,\rho';l_{\kappa})],
\label{eq:method_error_1}
\end{equation}
where the subscript ``$\kappa$'' can be either the cluster operator truncation ``cc'' or finite-size effect ``fs'', and
\begin{equation}
R_{\kappa}(\rho,\rho';l_{\kappa}) = y_{\kappa,\rm{ref}}(\rho)y_{\kappa,\rm{ref}}(\rho') r(\rho,\rho';l_{\kappa}). 
\label{eq:method_error_2}
\end{equation}
As before, we will use the corresponding Fermi momentum (in fm$^{-1}$) as the independent variable rather than the density.

For the cluster operator truncation we estimate the density-dependent mean error and covariance using results from a previous convergence study with 34 $\Delta$-full interactions at \NNLO{}~\cite{Hu:2021trw} (with the same $394$~MeV momentum cutoff as here). In that study, computations were performed at CCD(T) level which is a more accurate, but computationally heavier, \CC{} approximation that includes doubles excitations and perturbative triples corrections~\cite{hagen2013b}. In particular, triples correlation energies---the difference between CCD and CCD(T) results---were extracted for the 34 interactions.
We use the average (per density) triples correlation energy as our mean errors, resulting in $\mu_{\rm{cc,PNM}}(\rho) = 0.16 k_{\rm F}^2 -0.50k_{\rm F}+ 0.32 $~MeV/nucleon for \PNM{} and $\mu_{\rm{cc,SNM}}(\rho) = -1.28 k_{\rm F} + 0.80$~MeV/nucleon for \SNM{}. At $\rho = 0.16$~{fm}$^{-3}$ these mean shifts are $-0.04$~MeV/nucleon for \PNM{} and $-0.91$~MeV/nucleon for \SNM{}.
Furthermore, we also use these mean triples correlation energies as our reference scale $y_{\rm{cc,ref}}(\rho)$ which implies that $\bar{c}_{\rm cc}$ should be interpreted as the ratio between the cluster truncation error and the triples correlation energy.
Following previous \CC{} convergence studies~\cite{hagen2009b,Ekstrom:2022yea} and expert elicitation we conservatively assign $\bar {c}_{\rm{cc}} = 0.1$ for both \PNM{} and \SNM{} corresponding to $\pm 20\%$ of the triples correlation energy as a 95\% degree-of-belief error estimate of corrections beyond the triples approximation. 
We also assume that the observed density-dependence of the energy differences between the CCD and CCD(T) results for the 34 interactions in Ref.~\cite{Hu:2021trw} provides a relevant measure of the correlation structure of the \CC{} truncation method error. This data is therefore used to train the \GP[s] and the inferred correlation lengths (within the density range $\rho \in [0.12, 0.20]$~fm$^{-3}$) are $l_{\rm{cc,PNM}}= 0.50~\rm{fm}^{-1}$ and $l_{\rm{cc,SNM}}= 0.58~\rm{fm}^{-1}$. 

For the finite-size effect, we use the CCD ground-state energy as the reference scale and set correlation length $l_{\rm{cc,PNM}}= 0.50~\rm{fm}^{-1}$ and $l_{\rm{cc,SNM}}= 0.58~\rm{fm}^{-1}$. Following the study in Ref~\cite{hagen2013b}, we take $\pm 0.5\%$($\pm 4\%$) for \PNM{}(\SNM{}) as estimates of the $95\%$ credible intervals of finite-size errors for each system. This gives  $\bar {c}_{\rm{fs,PNM}} = 0.0025$ ($\bar {c}_{\rm{fs,SNM}} = 0.02$) for the nuclear-matter calculations in this work.

Finally, we also use the \GP{} error model described in Eqs.~\eqref{eq:method_error_1} and \eqref{eq:method_error_2} to incorporate the emulator error $\varepsilon_{\rm{emu}}(\rho)$ in Eq.~\eqref{eq:predicted_observables}. These \GP[s] were trained by the differences between emulator predictions and CCD results with the latter then used as reference scale. The training data are taken from the CCD computations and emulator predictions of the 34 interactions in Ref.~\cite{Hu:2021trw}. For the \SNM{} emulator error we found $l_{\rm emu,SNM}= 0.38~\rm{fm}^{-1}$ and we use $\mu_{\rm emu,SNM} = 0$ and $\bar {c}_{\rm emu,SNM} = 0.01$. We ignore the small emulator error for \PNM{} due to the high accuracy achieved by the \PNM{} emulator predictions.

\begin{figure}[!htb]
\includegraphics[width=0.95\linewidth]{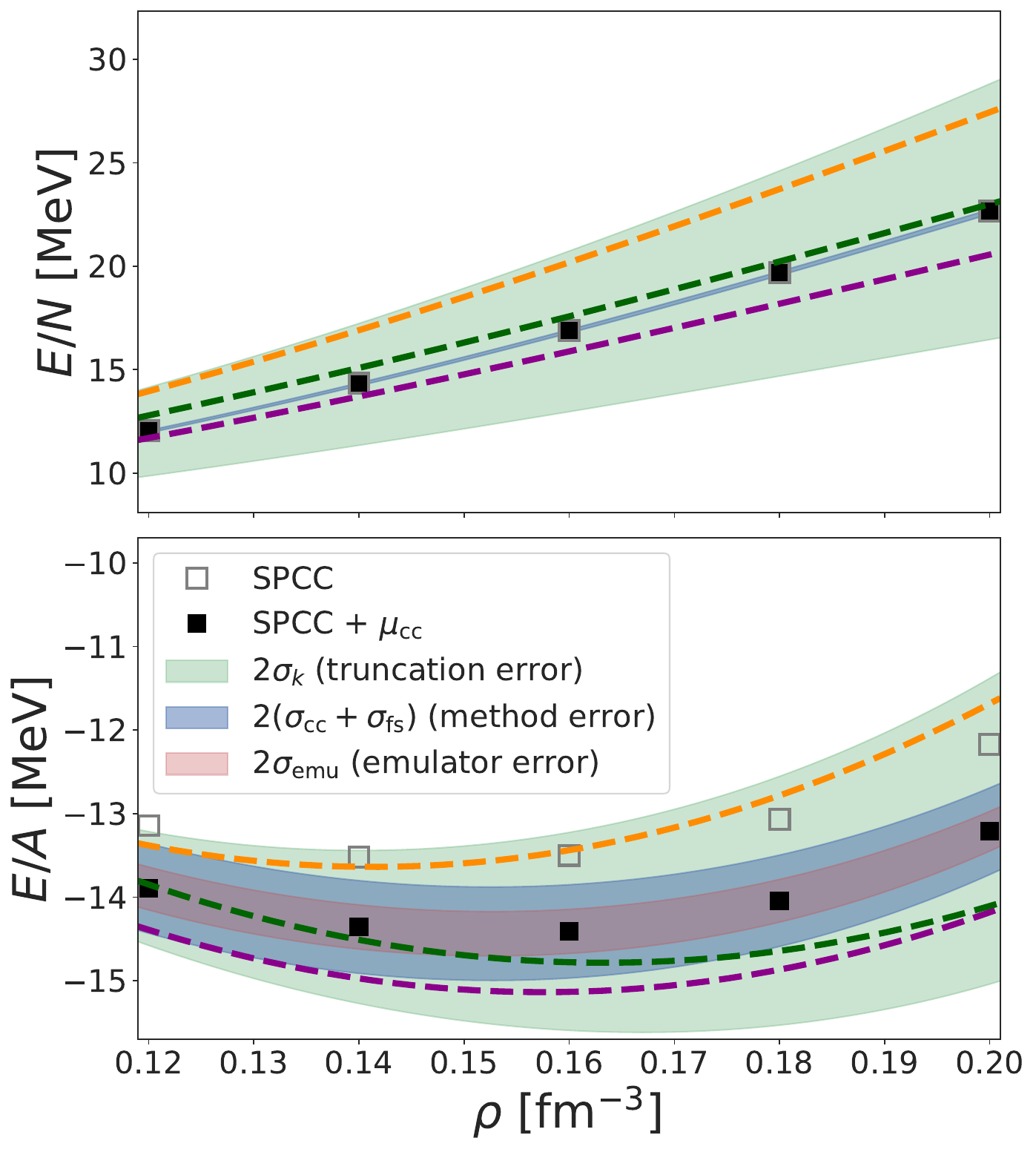}
\caption{(Color online) The \EOS{} for \PNM{} (top) and \SNM{} (bottom) calculated for one representative interaction with the nuclear matter emulators (open squares) plus the mean value of the method error (solid squares). The bands indicate two standard deviations of the truncation error (green), method error (blue) and emulator error (pink) from the \GP{} error models described in the text.
The errors at different densities are correlated as illustrated by three random samples shown by dashed curves. Correlations extend between \PNM{} and \SNM{} (sampled error curves in the same colour). 
\label{fig:EOS_with_error}
}
\end{figure}

Following these assignments, the full \PPD{} for nuclear matter observables, incorporating all relevant sources of uncertainty, can be sampled according to Eq.~\eqref{eq:predicted_observables}. In particular, it becomes straightforward to sample the error terms from the corresponding covariance matrices once the multitask \GP[s] are determined. In practice, this task is efficiently performed using
\begin{equation}
  \varepsilon = L x,
\label{eq:sample_}
\end{equation}
with $L$ being the Cholesky decomposition of the cross covariance matrix $K$ ($K= L L^{\rm{T}}$) and $x$ a standard normal random vector.
Note that we emulate results at five densities for both \PNM{} and \SNM{}. Thus $\varepsilon$ is a 10-dimensional vector and the cross covariance is a $10\times10$ matrix. This sampling procedure is crucial for generating the \PPD{} of nuclear matter properties.
The emulator predictions for the nuclear matter \EOS{} and the corresponding $2\sigma$ (95\%) credible interval for errors are illustrated in Fig.~\ref{fig:EOS_with_error}. Three randomly sampled \EOS{} predictions are also shown and one should note that the multitask \GP[s] guarantee that the sampled \EOS{} of \PNM{} and \SNM{} are smooth and properly correlated with each other. From this figure it is also clear that the method error for \PNM{} is quite small. This can be understood since emulator errors, finite-size effects and \CC{} correlation energies are all rather small for \PNM{}.

\section{History matching analysis}
The tremendous computational speed-up offered by our novel nuclear-matter emulators allows to perform a detailed statistical analysis of observable predictions using the \chiEFT{} model. The results shown in this section represent general outcomes of the interaction model described in Sec.~\ref{sec:method} used within \emph{ab initio} computations.

As a first step of this analysis we apply the history matching procedure as described in Sec.~\ref{section:HM} with five waves of global parameter search to iteratively reduce the \LEC{} domain. The history-matching is performed using neutron-proton phase shifts in $S$- and $P$-waves plus few-body ($A=2-4$) bound-state observables in the target sets.
The experimental values and the error assignments for the nuclear bound-state observables can be found in Table~\ref{tab:error_assignments}. The experimental targets are from Refs.~\cite{Wang:2020,angeli2013,machleidt2001}. Note that the target point-proton radii were transformed from experimental charge radii using the same relation as in Ref.~\cite{ekstrom2015a}. For the deuteron quadrupole moment we use the theoretical result obtained by the CD-Bonn~\cite{machleidt2001} model with a $4\%$ error bar. The theoretical model $\varepsilon_\mathrm{model}$ (method $\varepsilon_\mathrm{method}$) errors are estimated from the \EFT{} (\CC{}) convergence pattern as in Ref.~\cite{Hu:2021trw} while emulator errors $\varepsilon_\mathrm{em}$ are estimated from cross validation.
\begin{table}[!htb]
 \caption[Error assignments and PPD model checking]
 {Experimental values and error assignments for observables used in the fifth wave of the iterative history matching (history-matching observables) and for observables used in the model validation/calibration (predicted observables). Energies $E$ in (MeV), point-proton radii $r_p$ in (fm), and the deuteron quadrupole moment $Q$ in ($e^2\mathrm{fm}^2$). See the text for details.
   \label{tab:error_assignments}
}
\begin{center}
   \begin{tabular}{cccccc}
   \hline
   \multicolumn{6}{c}{History-matching observables}\\
   \hline
    Observable  & $z$ & $\varepsilon_\mathrm{exp}$ & $\varepsilon_\mathrm{model}$
    & $\varepsilon_\mathrm{method}$ & $\varepsilon_\mathrm{emu}$    \\
    \hline 
    $E(^{2}\mathrm{H})$ &  -2.2298   &     0.0    & 0.05 & 0.0005     & 0.001\%  \\
    %
    $r_p(^{2}\mathrm{H})$ & 1.976    &    0.0     & 0.005     & 0.0002 &0.0005\%   \\
    $Q(^{2}\mathrm{H})$ & 0.27      &      0.01   & 0.003 &   0.0005     &0.001\%   \\
    $E(^{3}\mathrm{H})$ &  -8.4818   &     0.0	  & 0.17	 & 0.0005         &0.01\%  \\
    $E(^{4}\mathrm{He})$ & -28.2956   &   0.0     & 0.55     &   0.0005       &0.01\%  \\
    $r_p(^{4}\mathrm{He})$ & 1.455  &    0.0     & 0.016    &  0.0002         &0.003\%   \\
   \hline
   \multicolumn{6}{c}{Predicted observables}\\
   \hline
    $E(^{6}\mathrm{Li})$ & -31.9940   &   0.0     & 0.55     &   0.2000       &0.01\%  \\
    $E(^{16}\mathrm{O})$ & -127.62   &   0.0     &  1.00    &   0.75        &0.5\%  \\
    $r_p(^{16}\mathrm{O})$ & 2.58  &    0     & 0.03    &  0.01        &0.5\%   \\
    \hline
  \end{tabular}
\end{center}
\end{table}

This selection of target data is representative of what could have been considered when seeking an optimal interaction model. However, the aim of our approach is fundamentally different. Rather than seeking a single optimum, we consider all non-implausible parametrizations in order to make a comprehensive study of the behaviour of our model. Furthermore, we consider much simpler linearised probability distributions, with just mean values and variances as specifiers, to identify the interesting parameter domain. Finally, we just divide the parameter space into implausible or non implausible. All samples from the latter domain are included in this part of the analysis without any probability weighting.

In the final wave, we explored $1\times 10^9$ samples from a space-filling design in the non-implausible domain that was established at the end of wave 4. We confronted the model predictions for the six $A=2-4$ observables and found $1.7\times10^6$ non-implausible interaction parametrizations.
At this point, we did not see any need to proceed with another wave since there were no signs of further reduction of the parameter domain. The $1.7\times10^6$ samples constitute a good representation of all non-implausible interactions.

\begin{figure}[!htb]
\includegraphics[width=1\linewidth]{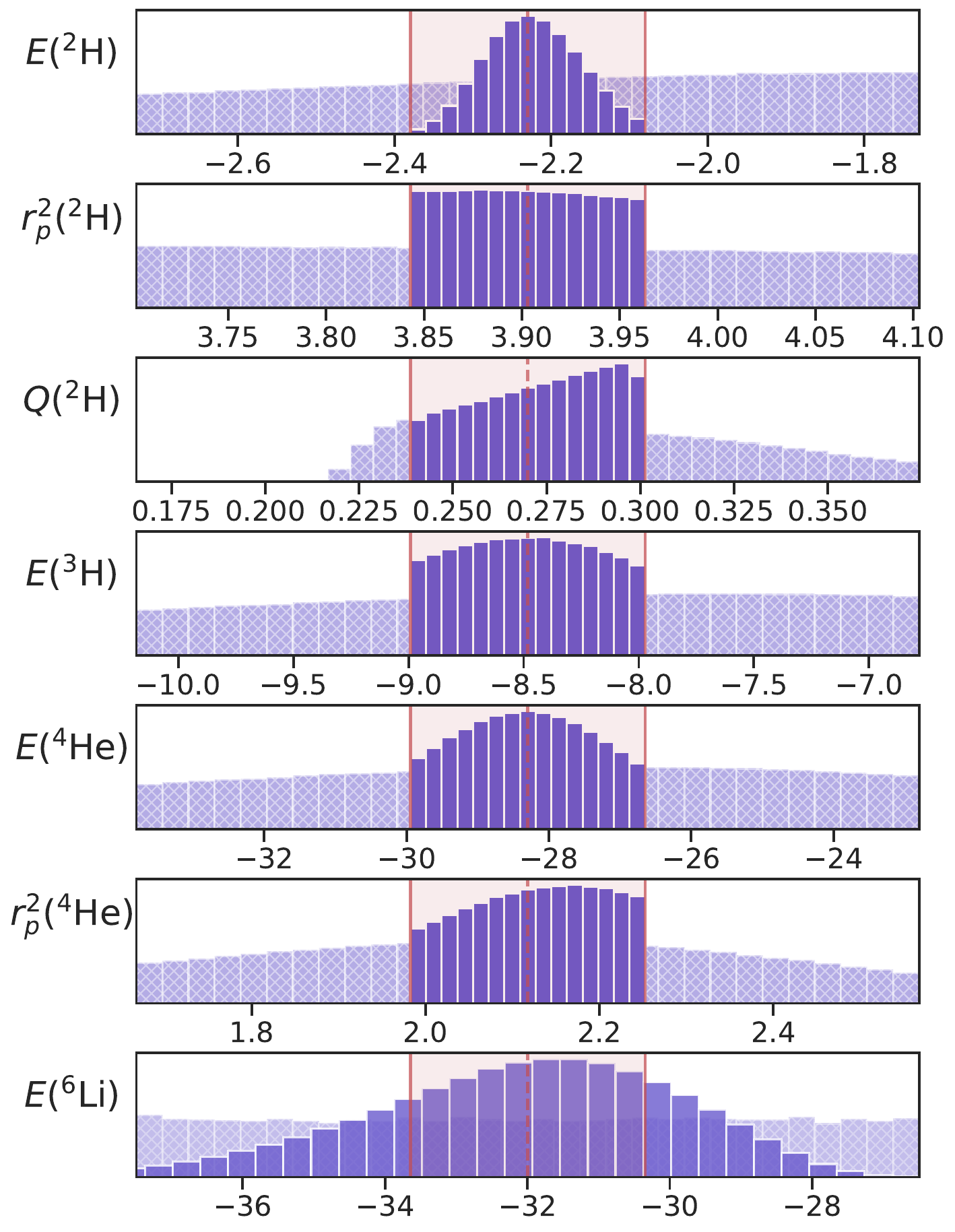}
  \caption{(Color online) Histograms of $A=2 -  6$ few-body observables predicted during history matching. The hashed histograms represent results obtained with $1\times 10^9$ random samples from the wave 5 input \LEC{} domain. The solid histograms correspond to model predictions with the final set of $1.7\times10^6$ non-implausible samples.
    Energies in (MeV), square of point-proton radii in ($\mathrm{fm}^2$) and the deuteron quadrupole moment in ($e^2\mathrm{fm}^2$). The red dashed lines denote the experimental values and the red bands indicate the $\pm 3\sigma$ error region.
\label{fig:hist_fewbody}
}
\end{figure}

Predictions for different few-body observables are shown in Fig.~\ref{fig:hist_fewbody}. Here we compare model predictions made with the $1\times 10^9$ random samples generated at the start of wave 5 (hashed histograms) with the results obtained with the $1.7\times10^6$ samples that survive the implausibility constraint.
As shown, the predictions with the random samples are characterized by very large variances. Clearly, the 17-dimensional \LEC{} domain is still quite large even after the history matching waves. As for the $1.7\times10^6$ non-implausible samples, all of them give results within $\pm 3\sigma$ error regions for the $A=2 - 4$ observables (as indicated by the red band) since these observables were included as target data in the history matching procedure and we used $c_I=3$ for the implausibility constraint~\eqref{maximum_implausibility}.

The prediction for the \nuc{6}{Li} ground state energy (last panel of Fig.~\ref{fig:hist_fewbody}) serves as model validation since it was not included in the history matching. We can see that the mode of the $E(\nuc{6}{Li} )$ histogram is reasonable, and clearly within the $3\sigma$ region, which indicates a very reasonable model performance for light nuclei.

\begin{figure}[!htb]
\includegraphics[width=0.90\linewidth]{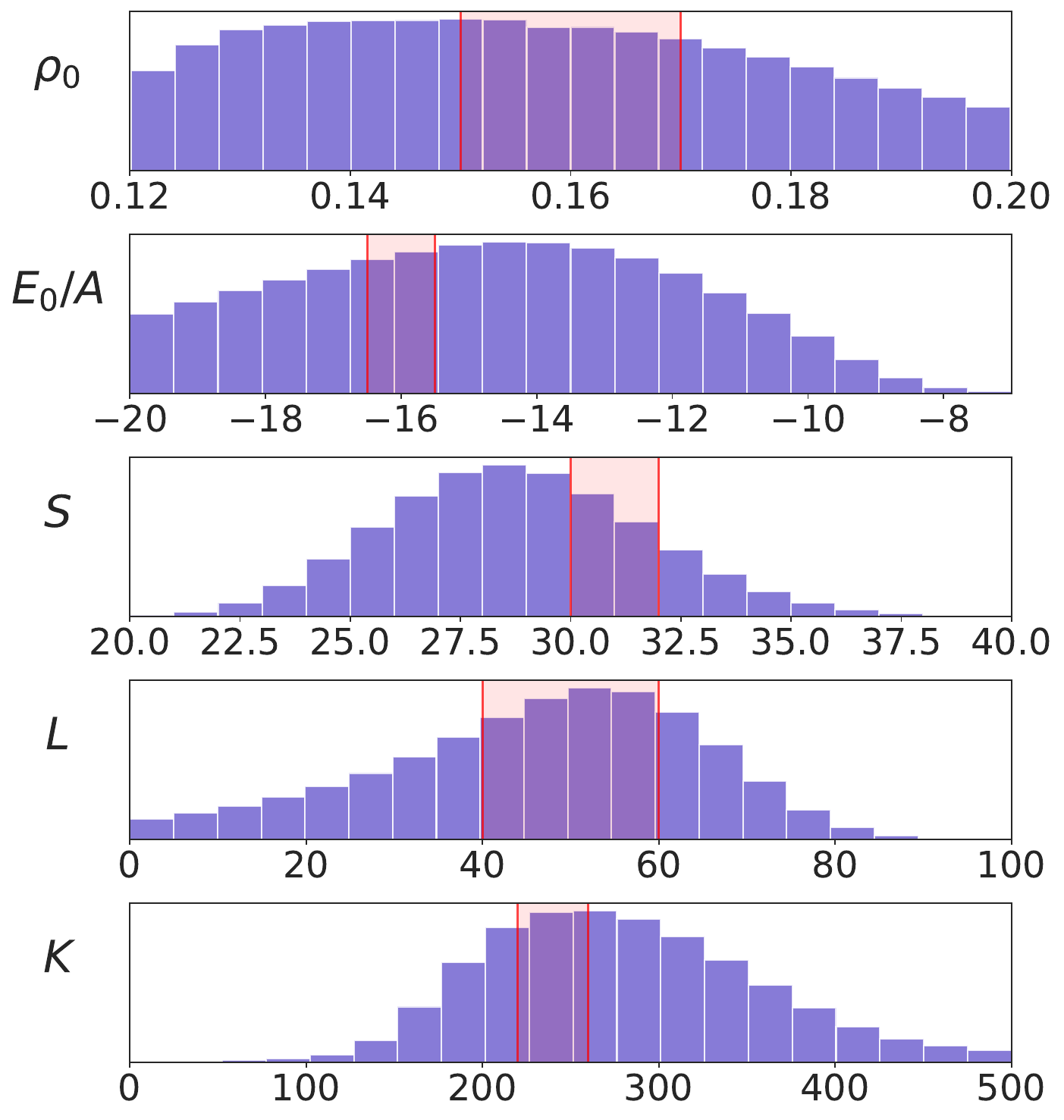}
\caption{(Color online) Histograms of nuclear matter properties at saturation predicted with the $1.7\times10^6$ non-implausible samples. Saturation density $\rho_0$ in ($\mathrm{fm}^{-3}$), saturation energy $E_{0}/A$, symmetry energy $S$, slope $L$ and incompressibility $K$ in ($\rm{MeV}$). The red bands indicate the empirical region with $E_0/A = -16.0 \pm 0.5$, $\rho_0 = 0.16 \pm 0.01$, $S=31 \pm 1$, $L=50\pm 10$ and $K=240\pm20$ from Refs.~\cite{lattimer2013,bender2003,shlomo2006}.
\label{fig:hist_NM}
}
\end{figure}

\begin{figure}[!htb]
\includegraphics[width=0.90\linewidth]{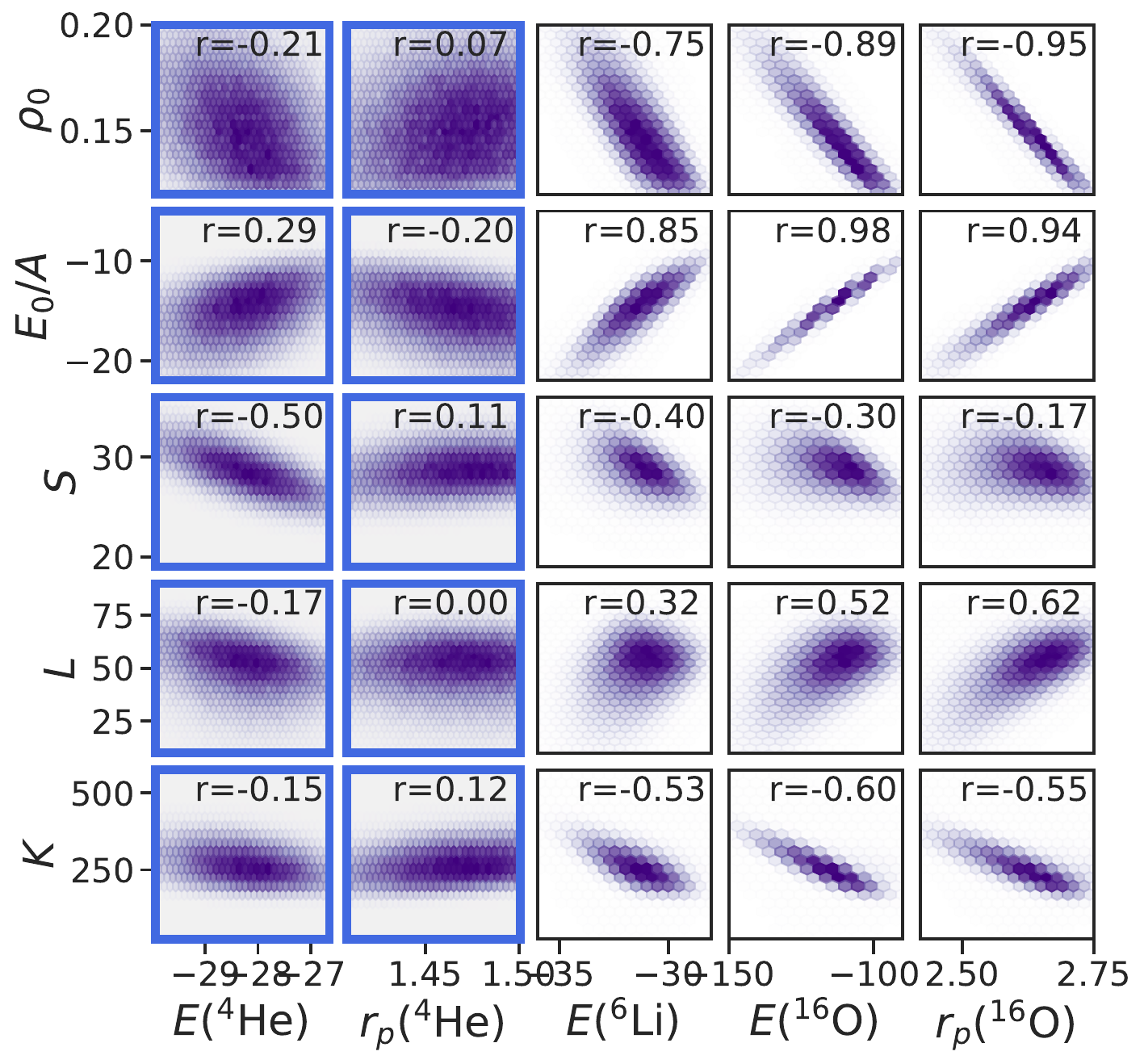}
\caption{(Color online) Correlation structure between nuclear matter saturation properties: saturation density $\rho_0$ (in fm$^{-3}$), saturation energy $E_{0}/A$, symmetry energy $S$, slope $L$, and incompressibility $K$ (all in MeV) and selected observables of  finite nuclei: ground-state energies (in MeV) and radii (in fm) of $^4{\rm He}$, $^6{\rm Li}$,$^{16}{\rm O}$. All results are obtained with the $1.7 \times 10^6$ non-implausible interactions from the fifth wave of history matching. The Pearson correlation coefficient $r$ is indicated in each panel. Note that $^4{\rm He}$ observables (thick, blue panel axes) are included in the history matching procedure while the other observables are pure predictions.
\label{fig:NM_vs_FN_HM}
}
\end{figure}

\begin{figure}[!htb]
\includegraphics[width=1\linewidth]{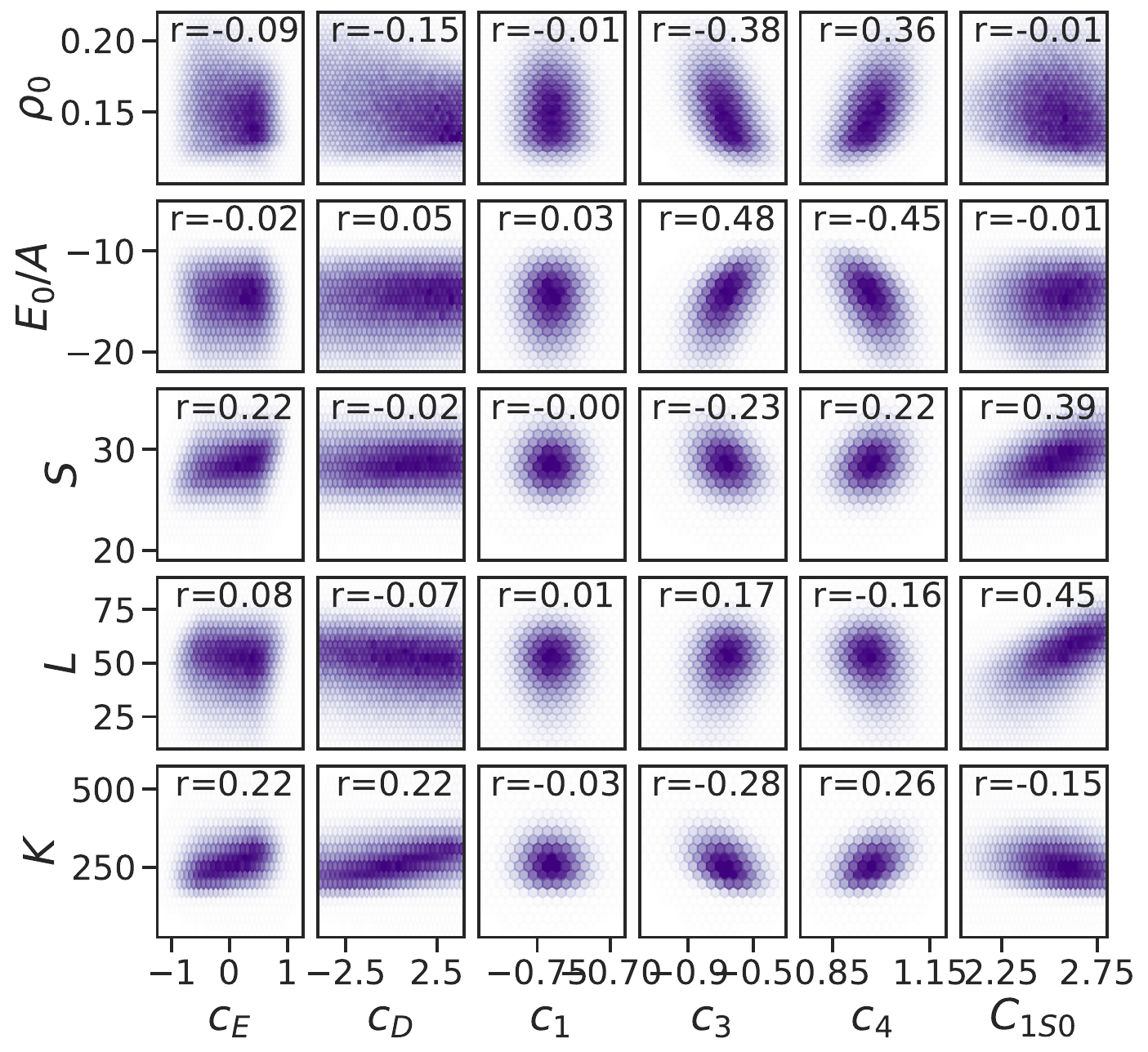}
  \caption{(Color online) Correlation structure between nuclear matter saturation properties: saturation density $\rho_0$ (in fm$^{-3}$), saturation energy $E_{0}/A$, symmetry energy $S$, slope $L$, and incompressibility $K$ (all in MeV) and selected \LEC[s] (\cD{}, \cE{}, $C_{1S0}$, and $C_{3P0}$). The parameters $c_i$ and $C_{1S0}$ are in units of $\text{GeV}^{-1}$ and $10^4$ GeV$^{-4}$, respectively. Results are obtained with the $1.7 \times 10^6$ non-implausible interactions from the fifth wave of history matching. The Pearson correlation coefficient $r$ is indicated in each panel.
\label{fig:NM_vs_LEC_HM}
}
\end{figure}

We then consider model predictions for the infinite nuclear matter systems using the \SPCC{} emulators from Sec.~\ref{sec:SPCC} with small-batch voting and \GP{} interpolation as described in Secs.~\ref{section:SBV} and \ref{sec:NMsaturation}.
In this particular analysis we don't perform a full sampling of the error model outlined in Sec.~\ref{section:error} but only include the mean shift of the \EOS{} for \SNM{}  that is expected from triples corrections. This shift is applied for all results shown in Figs.~\ref{fig:hist_NM}, \ref{fig:NM_vs_FN_HM}, and \ref{fig:NM_vs_LEC_HM}.

Saturation properties for the non-implausible interaction samples are shown in Fig.~\ref{fig:hist_NM}. All results are obtained using the nuclear matter emulator outputs as described in Sec.~\ref{sec:NMsaturation}. Interactions that give a saturation density outside of the interval $\rho \in [0.12, 0.20]$~fm$^{-3}$ (about 27\% of all non-implausible interactions) are not shown since our emulators are only constructed within this density interval. 
It is quite clear from Fig.~\ref{fig:hist_NM} that the modes for saturation density, saturation energy and the symmetry energy deviate from the empirical region and that here is a very large variance (in particular for the saturation density). We hypothesize that this is a consequence of the large extrapolation from the history matching observables in light nuclei ($A=2 -  4$) to properties of inifinite nuclear mater, and to the limited correlation (see below) between few-body observables and properties of nuclear matter.

The large number of non-implausible interaction samples together with access to fast and accurate emulators enable an extensive study of correlations between properties of finite nuclei and infinite nuclear matter obatined with chiral forces. The results of such a correlation study are shown in Fig.~\ref{fig:NM_vs_FN_HM}. Note that the $^6{\rm Li}$ and $^{16}{\rm O}$ observables are model predictions while the $^4{\rm He}$ ones are part of the history-matching procedure. We observe a positive correlation between ground-state energies of finite nuclei and the saturation energy $E_0/A$. This correlation is getting stronger from \nuc{4}{He} via \nuc{6}{Li} to \nuc{16}{O} ($r =0.29$, $0.85$, and $0.98$, respectively). This is reasonable since the central density of heavier system is closer to the density of nuclear matter at saturation.
On the other hand we find an anti-correlation between ground-state energies and the saturation density $\rho_0$. This negative correlation is also getting more prominent in heavier system. As for the \nuc{16}{O} radius we observe a similar correlation structure as for the energy meaning a positive correlation with $E_0/A$ and anti-correlation with $\rho_0$. We stress that these correlations are general results of the design of the $\Delta$NNLO interaction model and that they are characteristic features of the corresponding Hamiltonian.

The correlation between selected \LEC[s] and nuclear matter properties are shown in Fig.~\ref{fig:NM_vs_LEC_HM}. Even though \TNF[s] should be important for an accurate description of the saturation of nuclear matter, we find that the correlation between the 3NF short-range contacts \cD{} and \cE{} and nuclear matter properties is weak (as indicated by small Pearson correlation coefficients). This observation is consistent with the fact that \cD{} and \cE{} are not well constrained by the present history matching observables and that their contribution to the \TNF[s] are not exclusive since other \LEC[s] such as $c_{1,3,4}$ also play an important role.
It is interesting to note that the singlet $S$-wave contact $C_{1S0}$ gives the strongest correlation with the slope $L$ of the symmetry energy, which is also known to be strongly correlated with the neutron skin thickness of \nuc{208}{Pb}~\cite{brown2000,Hu:2021trw}. This result indicates that this particular \LEC{} serves as a bridge between neutron-proton scattering in the $^1S_0$ partial wave and the thickness of neutron skins in finite nuclei. A more detailed discussion of this constraint on the allowable range of the \nuc{48}{Ca} and \nuc{208}{Pb} neutron skin thicknesses can be found in \textcite{Hu:2021trw}. 

\section{Bayesian analysis: Posterior predictive distributions}
\label{section:Bayesian}
The history matching results do not offer a probabilistic interpretation since no actual probability distributions were invoked. For the target data we only considered an implausibility criterion rather than a fully specified likelihood. The non-implausible samples do, however, offer an excellent starting point for a Bayesian analysis.
In order to acquire \PPD[s] for nuclear matter and finite nuclei observables we proceed as follows: First, since the final wave of history matching procedure does not include phase shifts, we confront all $1.7\times10^6$ non-implausible samples with the phase shift targets from the first wave (including $S$ and $P$ partial waves up to $T_{\rm lab}=200$ MeV) and apply the implausibility constraint~\eqref{maximum_implausibility}.
We also examine whether the samples give an $np$ bound state in the $^1S_0$ channel as a sanity check. Just a few samples failed this test. Taken together, these constraints reduce the number of non-implausible samples to 8,218. Second, we use the method of sampling/importance resampling~\cite{smith:1992aa, Jiang:2022off} to extract an approximate posterior \PDF{} of the \LEC[s] via Bayes' theorem

\begin{equation}
  {\rm pr}(\vec{\alpha} \, | \, {\Dcal}) \propto \likelihood{\Dcal}{\vec{\alpha}} \, {\rm pr}(\vec{\alpha}).
\label{eq:pdf}
\end{equation} 
We assume a uniform prior probability distribution, ${\rm pr}(\vec{\alpha})$, for all \LEC[s] except $c_{1,2,3,4}$ for which the prior is  described by a multivariate normal distribution originating in the Roy-Steiner analysis of $\pi N$ scattering data performed in Ref.~\cite{siemens2017}. The history matching procedure provides a set of samples from this prior. Although the full data likelihood is not involved, the incorporation of implausibility constraints guarantees that samples with a negligible contribution to the posterior \PDF{} are removed. Operating with the remaining large set of prior samples $\{\vec{\alpha}_i\}_{i = 1}^n$ we now specify a data likelihood and evaluate $\omega_i \equiv \likelihood{\Dcal}{\vec{\alpha_i}}$ and so called importance weights $q_i \equiv \omega_i/\sum_{j=1}^n \omega_j$.
Finally, we resample a set $\{ \vec{\alpha}_i^* \}_{i=1}^N$ from the discrete distribution $\{\vec{\alpha}_i\}_{i = 1}^n$ according to the importance weights $q_i$. This resampled set will then be approximately distributed according to the target distribution $\prCond{\vec{\alpha}}{\Dcal} \propto \likelihood{\Dcal}{\vec{\alpha}} {\rm pr}(\vec{\alpha})$. See Ref.~\cite{Jiang:2022off} for a recent importance resampling review with a nuclear theory perspective. We have studied the convergence of the posterior and found that a resampling set of size $N=10,000$ is sufficient. We also found that a rather large subset of $>400$ samples from $\{\vec{\alpha}_i\}_{i = 1}^n$ provide 95\% of the posterior \PDF{} samples.

In order to examine how the choice of calibration data in the likelihood $\likelihood{\Dcal}{\vec{\alpha}}$ affects the \chiEFT{} prediction we considered two different versions: (i) $\Dcal = \Dfew$ encompassing binding energies and radii of \nuc{2,3}{H} and \nuc{4}{He} including the quadrupole moment of the deuteron, and (ii) $\Dcal = \Dmany$ where we also include the energy and radius of \nuc{16}{O}. 
The default choice for the functional form of the likelihood is a normal distribution with independent errors (as summarized in Table~\ref{tab:error_assignments}). The sensitivity to this specification of uncorrelated, Gaussian errors was tested using two alternative likelihood forms, namely a non-correlated Student-t distribution (with $\nu=5$ degrees of freedom implying heavier tails) and a multivariate normal distribution with positive correlation ($\rho=0.7$) between the ground-state energy and radius of the same nucleus and between ground-state energies of different nuclei.
In the end, we found no significant impact using the alternative distributions and therefore only show results obtained with the default, uncorrelated Gaussian likelihoods.
See the Supplemental material~\cite{supp2022:PRC} for numerical \LEC{} values of the non-implausible samples, the corresponding observable predictions using the emulators described in the text, and the likelihood \PDF{} values.

The model \PPD{} for an observable can be written as the set of model predictions evaluated for samples drawn from the parameter posterior
\begin{equation}
{\rm PPD_{\rm th}} = \{ \rm \textbf{y}_{th}(\vec{\alpha}) : \vec{\alpha} \sim {\rm pr}(\vec{\alpha} \, | \, {\Dcal})  \},
\label{eq:ppd}
\end{equation}
for which we use the resampled set $\{ \vec{\alpha}_i^* \}_{i=1}^N$.
From Eq.~\ref{eq:ppd} it is clear that the predictive distribution is conditional on the selected calibration data $\Dcal$.

\begin{figure}[!htb]
\includegraphics[width=0.90\linewidth]{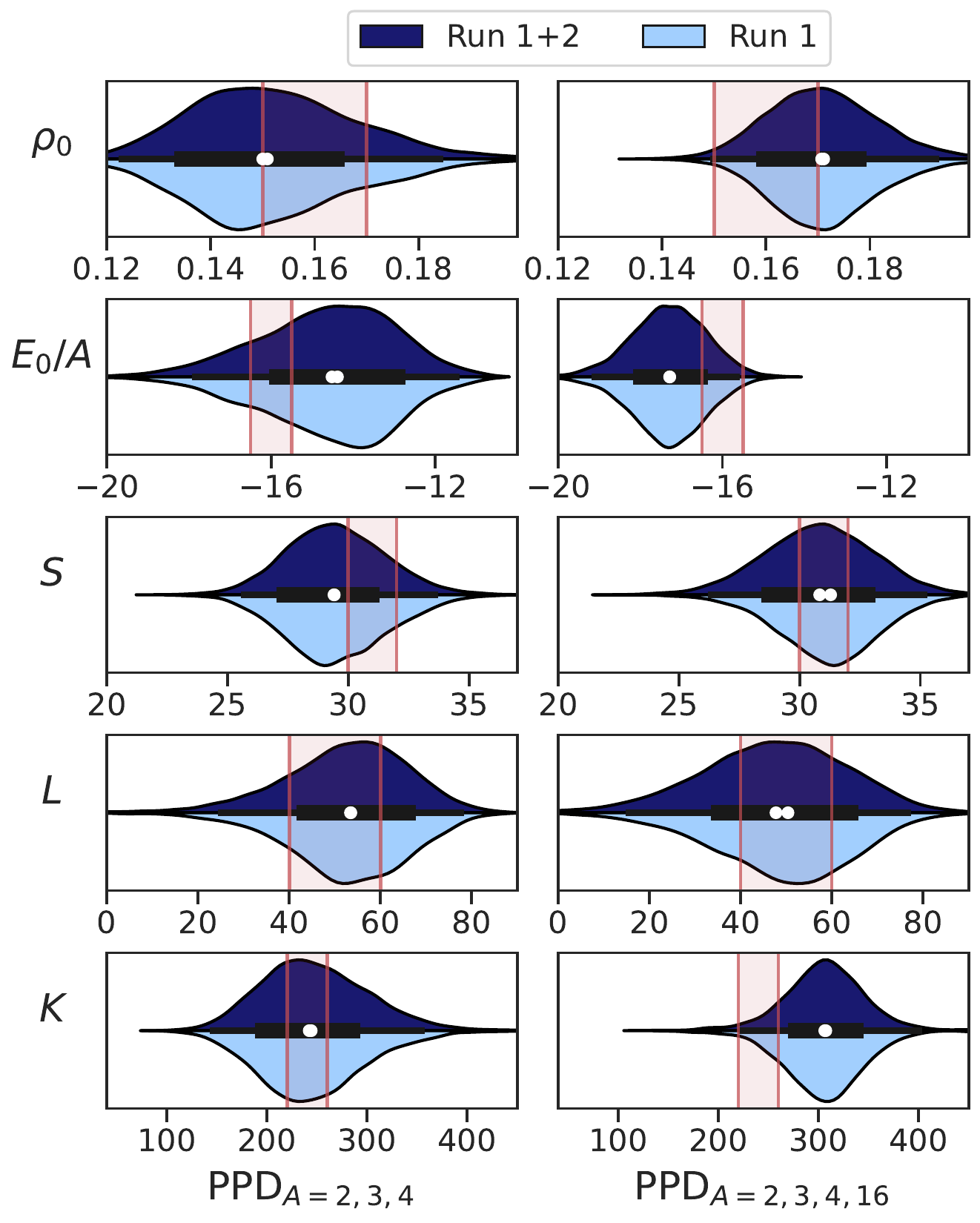}
\caption{(Color online) The \PPD{} of nuclear matter properties at saturation with two different choices of calibration observables: $\Dfew$ (left column) and $\Dmany$ (right column). The panels (from top to bottom) show the saturation density $\rho_0$ (in $\mathrm{fm}^{-3}$), saturation energy $E_{0}/A$, symmetry energy $S$, slope $L$, and incompressibility $K$ (all in $\rm{MeV}$). The red bands indicate the empirical region with $E_0/A = -16.0 \pm 0.5$, $\rho_0 = 0.16 \pm 0.01$, $S=31 \pm 1$, $L=50\pm 10$ and $K=240\pm20$ from Refs.~\cite{lattimer2013,bender2003,shlomo2006}. The upper half of each panel (dark blue) indicates \PPD{} results from the sum of two independent runs ($\approx$ 10,000 non-implausible samples in total) while the lower half (light blue) shows results obtained by the first run only($\approx$ 5,000 non-implausible samples).
\label{fig:ppd}
}
\end{figure}

\begin{figure}[!htb]
\includegraphics[width=0.95\columnwidth]{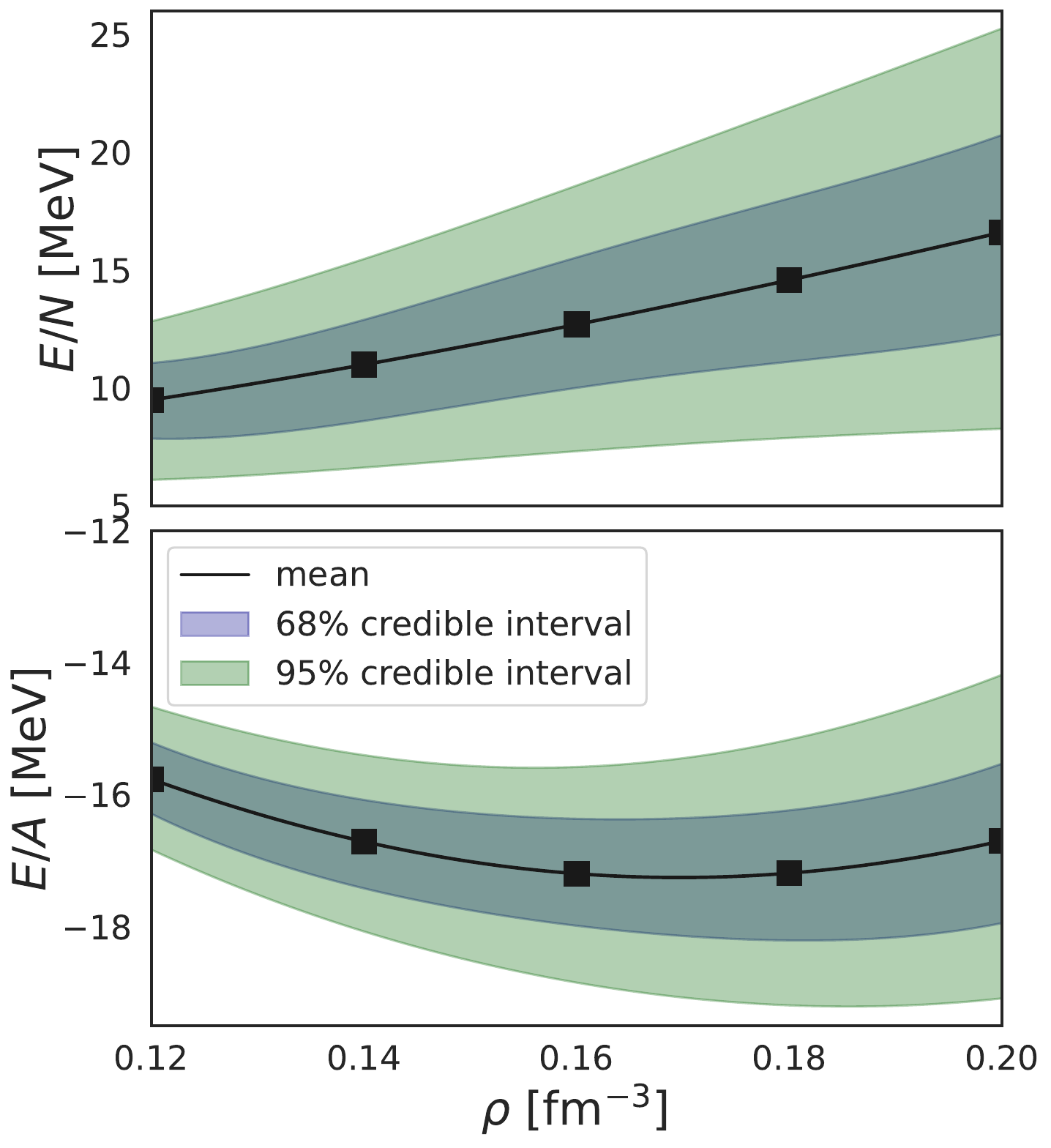}
\caption{(Color online) The \PPD{} for the EOS around saturation density, $\prCond{E(\rho)/N, E(\rho)/A}{\Dmany}$. The sampling of the \PPD{} includes all relevant errors as described in Sec.~\ref{section:error} as well as the parametric uncertainty.
\label{fig:EOS}
}
\end{figure}

Fig.~\ref{fig:ppd} shows the predicted distribution of nuclear matter properties calibrated by either $\Dfew$ or $\Dmany$.
Here we collect samples from the full \PPD{} for which we also sample different sources of uncertainty as discussed in Sec.~\ref{section:error}. For \ppdfew{}(\ppdmany{}) we find that $15\%(60\%)$ of the samples are drawn from the training set and therefore have no emulator error.
The \ppdfew{} is shown in the left column of Fig.~\ref{fig:ppd}. The modes of the marginal distributions for saturation density, saturation energy and symmetry energy still deviate from the empirical values. We note that the saturation density \PPD{} is asymmetric and our result almost indicates a bimodal distribution.
The \ppdmany{} is shown in the right column of Fig.~\ref{fig:ppd} and provides an improved prediction of nuclear matter saturation with better precision. The saturation energy is slightly lower (more binding) compared with the empirical range and the mode of the incompressibility $K$ is shifted to larger values. The predictions for $S$ and $L$ are not significantly affected by the addition of \nuc{16}{O} to the calibration data.
%
The comparison of \ppdfew{} and \ppdmany{} reveals that the description of nuclear matter properties is quite sensitive to the choice of calibration observables. The reason is quite clear from the correlation results shown in Fig.~\ref{fig:NM_vs_FN_HM}. It is clear that the \nuc{16}{O} ground-state energy and radii are strongly correlated with nuclear matter properties. Thus a likelihood that contains these observables provides a more precise nuclear matter prediction.
The full \PPD{} for the energy per particle of \PNM{} (top panel) and \SNM{} (bottom panel) as a function of density is shown in Fig.~\ref{fig:EOS}.

Given the multi-stage analysis one can ask whether the final result is sensitive to the randomness of non-implausible samples that results from the space-filling designs used in history matching.  For this reason we performed two independent runs, from the start of the history matching to the final Bayesian analysis, with each one producing $\approx$ 5,000 non-implausible samples. It is the sum of those two runs that is presented in the upper half of each panel in Fig.~\ref{fig:ppd}. The lower half displays the smaller statistics result that is obtained with just the first run. The similarity of the upper and lower halves indicates the robustness of the approach and the fact that the convergence of the sampling/importance resampling step is sufficient to accurately represent the target distribution.

\begin{figure}[!htb]
\includegraphics[width=0.90\linewidth]{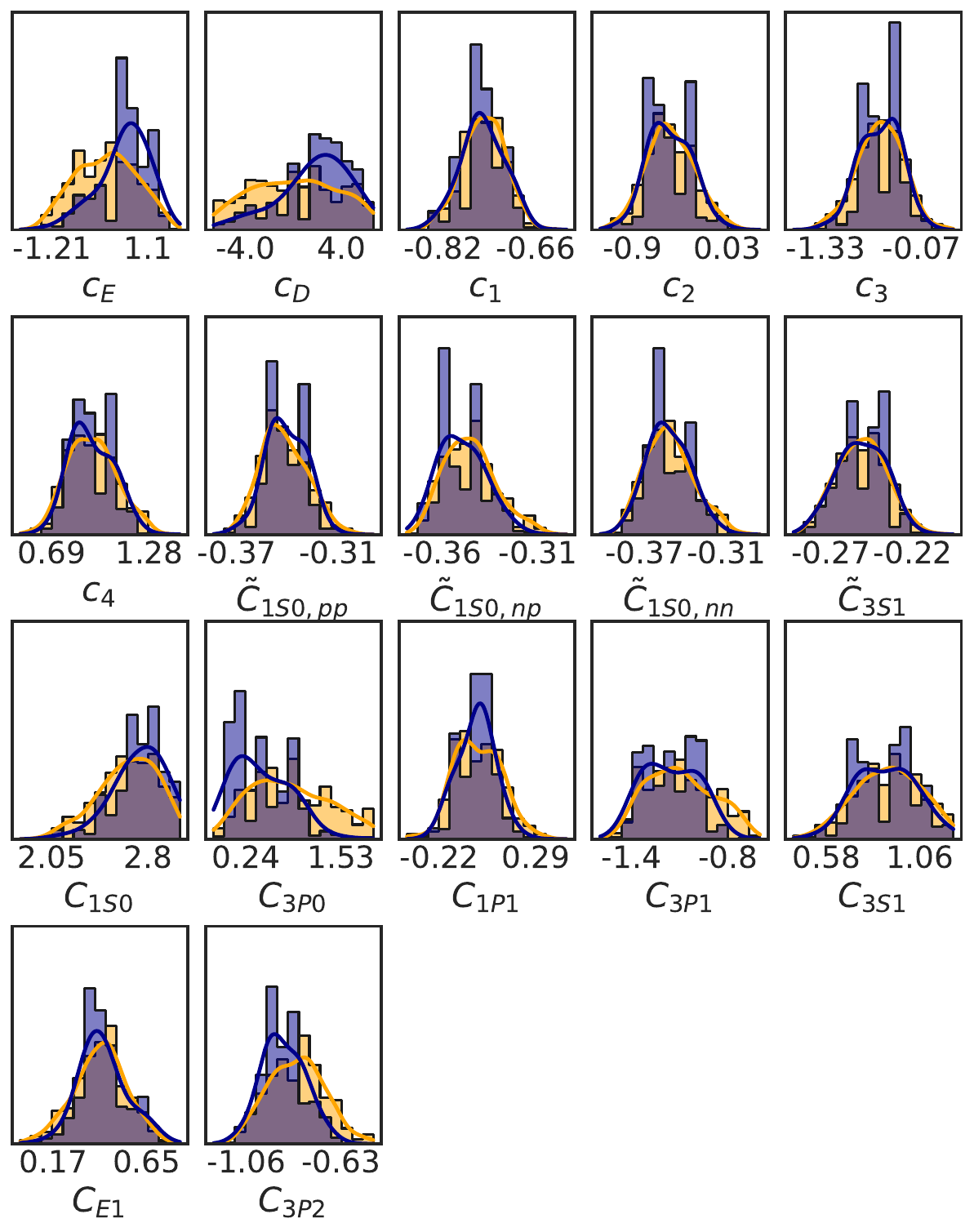}
\caption{(Color online) The \LEC{} posterior \PDF{} from importance resampling with either $\Dfew$ (orange) or $\Dmany$ (blue) as calibration data. The range for each \LEC{} is listed below each panel. The parameters $c_i$, $\tilde{C_i}$ and $C_i$ are in units of $\text{GeV}^{-1}$, $10^4$ $\text{GeV}^{-2}$ and $10^4$ $\text{GeV}^{-4}$, respectively.
\label{fig:lec_hist_2_pdf}
}
\end{figure}

Finally, in Fig.~\ref{fig:lec_hist_2_pdf} the \LEC{} parameter \PDF{} is shown conditional on the two calibration data sets. The marginal distribution of $c_D$, $c_E$, $C_{3P0}$, $C_{3P1}$ and $C_{3P2}$ are the most sensitive ones with respect to the choice of calibration data, while the differences in the other marginal distributions are barely distinguishable. This finding suggests that these terms in the chiral Hamiltonian are the most important ones to describe nuclear matter and heavier mass nuclei---yet remains poorly constrained by observables in the few-nucleon sector.

\section{Summary}
We have constructed nuclear matter emulators using the \SPCC{} method that works for a large 17-dimensional \LEC{} hyperspace of $\Delta$-full \chiEFT{} at \NNLO{}. In particular, we have developed a small-batch voting algorithm to handle the spurious-state problem that can occur when emulating quantum many-body methods employing a non-Hermitian Hamiltonian.
These nuclear matter emulators are then applied to $1.7 \times 10^6$ non-implausible interaction samples generated via five waves of history matching with $A=2-4$ observables. This allows to study properties of the \chiEFT{} model including the correlation structure between nuclear matter saturation properties and observables of finite nuclei without bias from a specific optimization scheme.
In particular we find an increasing correlation between saturation energy/density and the ground-state energy/radius of finite nuclei as the mass number of nuclei increases. In addition, a positive correlation between $C_{1S0}$ and the symmetry energy slope $L$ is observed.

Starting from the history matching samples we performed a Bayesian analysis including relevant sources of uncertainty and using a correlated error model for the nuclear \EOS{}. We applied the method of sampling/importance resampling method to obtain approximate samples of two parameter posterior \PDF{}s with two different calibration data sets, \Dfew{} and \Dmany{}.
The corresponding nuclear matter predictions (given by \ppdfew{} and \ppdmany{}) illustrate the sensitivity to the calibration data. We found that predictions of nuclear matter saturation is more precise when incorporating the \nuc{16}{O} energy and radius in the likelihood calibration.

We conclude that observables from \nuc{16}{O} are informative, but we note that they are not the only choice. We have seen that predictions for $S$ and $L$ were not significantly affected by the addition of \nuc{16}{O} to the calibration data. It will therefore be interesting to explore the information content of observables from neutron-rich systems.
Furthermore, one should also consider other few-nucleon observables, such as the $^{3}$H $\beta$ decay rate for which emulators are also available~\cite{Wesolowski:2021cni}, to monitor how they constrain the chiral interaction model and to explore whether this can lead to a satisfactory description of nucleonic matter ranging from light nuclei to infinite nuclear matter. 

\begin{acknowledgements}
  We thank Andreas Ekstr\"om and Thomas Papenbrock for useful discussions. This work was supported by the Swedish Research Council (Grant Nos 2017-04234 and 2021-04507), the European Research Council under the European Unions Horizon 2020 research and innovation program (Grant No. 758027), and the U.S. Department of Energy under contract DE-AC05-00OR22725 with UT-Battelle, LLC (Oak Ridge National Laboratory).
  The computations and data handling were enabled by resources provided by the Swedish National Infrastructure for Computing (SNIC) at Chalmers Centre for Computational Science and Engineering (C3SE), and the National Supercomputer Centre (NSC) partially funded by the Swedish Research Council through Grant No. 2018-05973.
\end{acknowledgements}

\bibliography{./master,./temp}
\end{document}

%% file: shorthands.tex
\newcommand{\prCond}[2]{\ensuremath{\mathrm{pr}\left(#1 \, \vert \, #2 \right)}}
\newcommand{\likelihood}[2]{\ensuremath{\mathcal{L}\left(#1 \, \vert \, #2 \right)}}
\newcommand{\Dcal}{\ensuremath{\mathcal{D}_\mathrm{cal}}}
\newcommand{\Dfew}{\ensuremath{\mathcal{D}_{A=2,3,4}}}
\newcommand{\Dmany}{\ensuremath{\mathcal{D}_{A=2,3,4,16}}}
\newcommand{\ppdfew}{\ensuremath{{\rm PPD}_{A=2,3,4}}}
\newcommand{\ppdmany}{\ensuremath{{\rm PPD}_{A=2,3,4,16}}}

\def\nuc#1#2{\relax\ifmmode{}^{#1}{\protect\text{#2}}\else${}^{#1}$#2\fi}
\def\itnuc#1#2{\setbox\@tempboxa=\hbox{\scriptsize\it #1}
\def\@tempa{{}^{\box\@tempboxa}\!\protect\text{\it #2}}\relax
\ifmmode \@tempa \else $\@tempa$\fi}

\newcommand{\newabbreviation}[3]{\newcounter{#1}\expandafter\newcommand\csname#1\endcsname[1][]{\ifthenelse{\equal{##1}{abreviate}}{#2}{\ifthenelse{\equal{##1}{fullname}}{#3}{\ifthenelse{\equal{##1}{explain}}{#3 (#2)\stepcounter{#1}}{\ifthenelse{\value{#1}=0}{#3##1 (#2##1)\stepcounter{#1}}{#2##1}}}}}}

\newabbreviation{EFT}{EFT}{effective field theory}
\newabbreviation{chiEFT}{$\chi$EFT}{chiral effective field theory}
\newabbreviation{LO}{\text{LO}}{leading order}
\newabbreviation{NLO}{\text{NLO}}{next-to-leading order}
\newabbreviation{NNLO}{\text{NNLO}}{next-to-next-to-leading order}
\newabbreviation{LEC}{\text{LEC}}{low-energy constant}
\newcommand{\cD}{\ensuremath{c_{D}}}
\newcommand{\cE}{\ensuremath{c_{E}}}

\newabbreviation{EC}{\text{EC}}{eigenvector continuation}
\newabbreviation{CC}{\text{CC}}{coupled-cluster}
\newabbreviation{NCSM}{\text{NCSM}}{no-core shell model}
\newabbreviation{SPCC}{\text{SPCC}}{subspace-projected coupled cluster}
\newabbreviation{PPD}{\text{PPD}}{posterior predictive distribution}
\newabbreviation{PDF}{\text{PDF}}{probability density function}
\newabbreviation{GP}{\text{GP}}{Gaussian processe}
\newabbreviation{NM}{\text{NM}}{nuclear matter}
\newabbreviation{PNM}{\text{PNM}}{pure neutron matter}
\newabbreviation{SNM}{\text{SNM}}{symmetric nuclear matter}
\newabbreviation{EOS}{\text{EOS}}{equation-of-state}

\newabbreviation{NN}{\text{NN}}{nucleon-nucleon}
\newabbreviation{TNF}{\text{3NF}}{three-nucleon force}

